\documentclass[review]{elsarticle}
 \biboptions{comma,sort&compress}
\usepackage{graphicx}
\usepackage{amsmath}
\usepackage{here}
\usepackage{cuted}
\def\beq{\begin{equation}}
\def\eeq{\end{equation}}
\def\beqn{\begin{equation*}}
\def\eeqn{\end{equation*}}
\def\bea{\begin{eqnarray}}
\def\eea{\end{eqnarray}}
\def\bq{\begin{quote}}
\def\eq{\end{quote}}

\def\nnb{\nonumber}
\def\ga{\left(}
\def\dr{\right)}

\def\lrar{\longrightarrow}
	
\def\Lrar{\Longrightarrow}
																
\def\nnb{\nonumber}

\def\la{\langle}
\def\ra{\rangle}

\def\ba{\vspace*{-0.2cm}\begin{array}}
\def\ea{\end{array}\vspace*{-0.2cm}}

\def\b{$\bullet~$}

\def\d{$\diamond~$}

\def\als{\alpha_s}

\def\gg{\la\alpha_s G^2 \ra}
\def\ggg{\la g^3f_{abc} G^aG^bG^c \ra}
\def\gggg{\la\als^2 G^4\ra}

\usepackage{amsmath}
\usepackage{slashed}
\usepackage{color}

\begin{document}

%\markboth{Stephan Narison, Montpellier (FR)}{ }
\begin{frontmatter}

%%
%%%%%%%%%%%%%%%%%%%%%%%%%%%%%%%%%%%%%%%%%%%%%%%%%
%\begin{document}
%\title{Fully heavy scalar molecules and  tetraquarks states from QCD at NLO}
%\title{On the Tetraquarks / Molecules Nature of X(2900) from QCD Laplace sum rule}
%\begin{center}
\title{ \large  %Beyond the One Resonance 
Slope of the topological charge, proton spin and the $0^{-+}$ pseudoscalar  di-gluonia spectra
%Puzzling Sigma and Scalar Gluonia
}
\author{Stephan Narison%\corref{cor1}
}
%\cortext[cor1]{ICTP-Trieste consultant for Madagascar.}
\address{Laboratoire
Univers et Particules de Montpellier (LUPM), CNRS-IN2P3, \\
Case 070, Place Eug\`ene
Bataillon, 34095 - Montpellier, France\\
and\\
Institute of High-Energy Physics of Madagascar (iHEPMAD)\\
University of Ankatso, Antananarivo 101, Madagascar}
\ead{snarison@yahoo.fr}

\begin{abstract}
\noindent
In this paper, we attempt to discuss the topics mentioned in the title by scrutinizing  and improving the  $0^{-+}$ pseudoscalar di-gluonia/glueballs sum rules within the standard SVZ-expansion at N2LO  {\it without instantons}. 

First, we reconsider the estimate of the slope of the topological charge $\sqrt{\chi'(0)}(Q^2=2\,\rm GeV^2)=24.3(3.4)$ MeV from low degree moments which imply $\int_0^1\, dx\, g_1^P(x) = 0.144(5)\,  [\rm data:~0.145(13)]$ for the first moment of the polarized proton structure function (proton spin) and $G_A^{(0)}(10~\rm GeV^2)\equiv\Delta u+\Delta d +\Delta s = 0.340(50)\,\,[\rm data=0.330(39)]$ for the singlet form factor of the axial current. 

Second, we work  with high degree moments and parametrize the spectral function beyond the  minimal duality ansatz: ``{\it One resonance $\oplus$ QCD continuum}" to get the $0^{-+}$ pseudoscalar  di-gluonia spectra.  Then, we obtain three groups of gluonia :  {\it The familiar light $\eta_1$} [singlet gluon component of the $\eta'(958)$] with $[M_{\eta_1},f_{\eta_1}]= [825(45), 905(72)]$ MeV which is important for understanding the $U(1)_A$ anomaly; {\it The two new medium gluonia} with $M_{P_{1a}}=1338(112)$ MeV and   $[M_{P_{1b}}=1462(117) $ MeV or their mean  $[M_{P_{1}},f_{P_{1}}]=[1397(81),594(144)]$ MeV which support the gluonium nature of the excellent experimental candidate $\eta(1405)$   and may bring a small gluon piece to the $\eta(1295)$; their corresponding 1st radial excitations\,: $M_{P'_{1a}}=1508(226)$ MeV and $M_{P'_{1b}}=1553(139)$ MeV with their mean: $[M_{P'_{1}},f_{P'_{1}}]=[1541(118),205(282)]$ MeV which may be identified (up to some eventual mixings with $\bar qq$ states) with the observed $\eta(1475,1700)$  states; {\it The heavy gluonium} with the mean mass: $[M_{P_2}, f_{P_2}]=[2751(140),500(43)]$ MeV which can be compared with the lattice results. 
One can remark the (natural) one to one correspondence between the pseudoscalar gluonia and their chiral scalar analogue from Ref.\,\cite{SNS21}: $\sigma(1)\to \eta_1;~G_1(1.55)\to P_{1}; ~[\sigma'(1.1),G'(1.56)]\to P'_{1a,1b};~G_2(3)\to P_2$ which is mainly due to the importance  of the QCD PT contributions in the sum rule analysis that are almost equal in these two channels. 
\end{abstract}
\begin{keyword} 

%QCD spectral sum rules, Perturbative and Non-Pertubative calculations,  Hadron and Quark masses, Gluon condensates 
%(11.55.Hx, 12.38.Lg, 13.20-Gd, 14.65.Dw, 14.65.Fy, 14.70.Dj)
{\footnotesize QCD Spectral Sum Rules; Perturbative and Non-perturbative QCD; Exotic hadrons; Masses and Decay constants.}
%(11.55.Hx, 12.38.Lg,.
\end{keyword}
%}
%\ccode{Pac numbers: 11.55.Hx, 12.38.Lg, 13.20-Gd, 14.65.Dw, 14.65.Fy, 14.70.Dj}  
\end{frontmatter}
%%%%%%%%%%%%%%%%%%%%%%%%%%%%%%%%%%
%\vspace*{-1.5cm}
\vfill\eject
\pagestyle{plain}
 %%%%%%%%%%%%%%%%%%%%%%%%%%%%%%%%%%%
% \section*{References}
%\end{document}

 \section{Introduction}
 %%%%%%%%%%%%%%%%%%%%%%%%%%%%%%%%%%%
Gluonia / Glueballs bound states are expected to be a consequence of QCD\,\cite{MIN}. However, despite considerable theoretical and experimental efforts, there are not (at present) any clear indication of their signature. The difficulty is also  due to the fact that the observed candidates can be a strong mixing of the gluonia with the $\bar qq$ light mesons or with some other exotic mesons (four-quark, hybrid states).

%In particular, the $(I=0)$ Isoscalar  Scalar Channel channel is overpopulated beyond the conventional scalar nonet such that one suspects that some of these states can be of exotic origins and may (for instance) contain some large gluon component in their wave functions. 
%%%%%%%%%%%%%%%%%%%%%%%%%%%%%%%%
\subsection*{\b Experimental facts of the pseudoscalar states below 2 GeV}
%%%%%%%%%%%%%%%%%%%%%%%%%%%%%%%%
There are experimental observation of $0^{-+}~\eta$-like mesons below 2 GeV\,\cite{PDG,AMSLER}:
\beq
\eta(1295),~~~~~~~~~~~\eta(1405),~~~~~~~~~~~\eta(1475),~~~~~~~~~~~\eta(1760)~,
\eeq
 where some of them can be glueball candidates.  as they are produced in some gluonia rich channels like e.g. $J/\psi$ radiative decays, $\bar pp$ annihilation at rest.  The fact that the $\eta(1295),~\eta(1405)$ are seen to decay  into $\eta(\pi\pi)_{S-wave}$ where the $(\pi\pi)_{S-wave}$ may originate from the $\sigma/f_0(550)$ meson  can be understood with  the  $U(1)_A$ gluon vertex\,\cite{VENEZIA,SNG,SNS21}  $ \la \eta(1295)\vert \theta_\mu^\mu \vert \eta\ra 
$ where the $\sigma$ is the dilaton associated to the trace of the energy momentum-tensor $ \theta_\mu^\mu$. This feature can indicates the presence  of the gluon component into the wave function of these two states. The absence of the $\eta(1405)$ from $\gamma\gamma$ collisions may indicate that it has a larger gluon component than the $\eta(1295,1475)$ where the two latter are assumed\,\cite{AMSLER} to be the radial excitations of the $\eta$ and $\eta'$ while the $\eta(1405)$ is expected to be an excellent gluonium candidate. 
%%%%%%%%%%%%%%%%%%%%%
\subsection*{\b A short reminder of different theoretical predictions}
%%%%%%%%%%%%%%%%%%%%%

\d An earlier qualitative analysis of Novikov et al.\,\cite{NOVIKOV,NSVZ}  and a tachyonic gluon mass\,\cite{CNZa} based on some  hadron scale hierarchy arguments expects a high mass (pseudo)scalar gluonium.  These arguments seem to be supported by a quantitative analysis\,\cite{SNG,ASNER,ZHANG,FORKEL,ZHU} using QCD spectral sum rules (QSSR) \`a la SVZ\,\cite{SVZ,ZA,SNR,BELLa,BERTa,SNB5,SNB6,SNB1,SNB2,SNB3,SNB4} within one resonance including or not the $\eta'$ contribution and (or not) direct instanton contribution leading to a gluonium mass from $(2.05\pm 0.19)$\,\cite{SNG} to  2.7 GeV\,\cite{ZHANG}.  We shall comment these results later on. 

\d Lattice simulations with one resonance find a mass in the range  (2.15--2.72) GeV\,\cite{RAGO,MEYER,MATHIEU,CHEN}.

\d An holographic model provides a mass of about 2.1 GeV\,\cite{HOLO}. 

\d However, a recent analysis using inverse dispersive problem approach to the the spectral function predicts a lower mass  around 1.75 GeV\,\cite{HSIANG}, while the flux tube \,\cite{FADEEV}  and $\eta$-$\eta'$-$G$ mixing\,\cite{LIU} models predictions are around 1.4 GeV in agreement with the previous experimental expectations for the $\eta(1405)$. 

%%%%%%%%%%%%%%%%%%%%%
\subsection*{\b Projets}
%%%%%%%%%%%%%%%%%%%%%%
\d Motivated by the failure of some of the previous approaches including QSSR to explain the data below 2 GeV, we scrutinize previous QSSR results and attempt to improve them by parametrizing the spectral function with more than one resonance and by working with high degree moments. 

\d ``En passant", we revise our NLO estimate of the slope of the topological charge $\chi'(0)$ in the chiral limit\,\cite{SHORE} and study the effect of this N2LO result on the spin of the proton. 

 %%%%%%%%%%%%%%%%%%%%%%%%%%%%%%%%%%%%%
%\section{Pseudoscalar gluonium Laplace sum rules}
 %%%%%%%%%%%%%%%%%%%%%%%%%%%%%%%%%%%%%
%%%%%%%%%%%%%%%%%%%%%%%%%%%%%%%%%%%%%%%%%%%
\section{The QCD anatomy of the pseudoscalar gluonium two-point correlator}
%%%%%%%%%%%%%%%%%%%%%%%%%%%%%%%%%%%%%%%%%%%
We shall work with  two-point correlator:
\beq
\psi_P(q^2) =\ga{8\pi}\dr^2i\int d^4x \,e^{iqx}\la 0 \vert Q(x)Q^\dagger(0)\vert 0\ra
\eeq
 associated to the  divergence of the $U(1)_A$ axial current  which reads for $n_f$ quark flavours ;
 \beq
 \partial_\mu J^\mu_5(x)= \sum_{i=u,d,s}2m_i\bar\psi_i\psi_i+2n_f Q(x)~, 
 \eeq
 where :
 \beq
Q(x)\equiv
 \ga\frac{\alpha_s}{16\pi}\dr\epsilon_{\mu\nu\rho\sigma}G^{\mu\nu}_a(x)G^{\rho\sigma}_a(x)
 \eeq
 is the topological charge density,  $a=1,...,8$ is the colour index,  $\alpha_s\equiv g^2/4\pi$ is the QCD coupling and $G^{\mu\nu}_a(x)\equiv \partial^\mu A_a^\nu - \partial^\nu A_a^\mu  + gf_{abc}A^{\mu,b}A^{\nu,c}$ is the Yang-Mills field strength constructed from the gluon fields\,$A^\mu_a$. 
%%%%%%%%%%%%%%%%%%%%%%%%%%%%%%%%%%%%%%%%%%%%%%%
\subsection*{\b The standard SVZ-expansion}
%%%%%%%%%%%%%%%%%%%%%%%%%%%%%%%%%%%%%%%%%%%%%%%
Using the Operator Product Expansion (OPE) \`a la SVZ , its QCD expression can be written as :

\beq
\psi_P(q^2)=2\sum_{0,1,2,...}C_{2n}\la  {\cal O}_{2n}\ra~.
\eeq
where $C_{2n}$ is the Wilson coefficients calculable perturbatively while $ \la  {\cal O}_{2n}\ra$ is a short-hand notation for the non-perturbative vacuum condensates $\la 0 \vert {\cal O}_{2n}\vert 0\ra$  of dimension $2n$.
%%%%%%%%%%%%%%%%%%%%%%%%%%%%%%%%%%%
\subsubsection*{\d The unit perturbative operator ($n=0$) }
%%%%%%%%%%%%%%%%%%%%%%%%%%%%%%%%%%%
Its contribution reads :
\bea
C_0&\equiv&- Q^4 L_\mu\Big{[} C_{00} +C_{01} L_\mu+C_{02} L_\mu^2\Big{]}~~{\rm with} :\nnb\\
C_{00} &=& a_s^2\ga 1+20.75a_s+305.95a_s^2\dr~,~~~C_{01}=-a_s^3\ga \frac{9}{4}+72.531a_s\dr~,~~C_{02}=5.0625a_s^4,
\label{eq:pert}
\eea
where the NLO (resp. N2LO) contributions have been obtained in\,\cite{KATAEV} (resp.\,\cite{CHETG,ZHANG}). $L_\mu\equiv {\rm Log}(Q^2/\mu^2)$ where $\mu$ is the subtraction point and $a_s\equiv \alpha_s/\pi$. We shall use for 3 flavours :
\beq
\Lambda = 340(28)~{\rm MeV},
\eeq
deduced from $\alpha_s(M_Z)= 0.1182(19)$ from $M_{\chi_{c0,b0}}-M_{\eta_c,\eta_b}$mass-splittings\,\cite{SNparam,SNparama},\,$\tau$-decays\,\cite{PICHTAU,SNTAU} and the world average\,\cite{PDG,BETHKE}. We shall use the running QCD coupling to order $\alpha_s^2$:
\beq
a_s(\mu)= a_s^{(0)}\Bigg{\{} 1-a_s^{(0)}\frac{\beta_2}{\beta_1}LL_\mu+\ga a_s^{(0)}\dr^2\Bigg{[}\ga\frac{\beta_2}{\beta_1}\dr^2\ga LL_\mu^2-LL_\mu-1\dr+\frac{\beta_3}{\beta_1}\Bigg{]}\Bigg{\}}~,
\label{eq:alphas}
\eeq
where :  
\beq
 a_s^{(0)}\equiv \frac{1}{-\beta_1{\rm Log} \ga{\mu}/{\Lambda}\dr}~~~~~~~{\rm and}~~~~~~~LL_\mu\equiv {\rm Log} \, \Big{[} {\rm 2\,Log}\ga{\mu}/{\Lambda}\dr\Big{]}. 
 \eeq
%%%%%%%%%%%%%%%%%%%%%%%%%%%%%%%%%%% 
\subsubsection*{\d The dimension-four gluon condensate ($n=2$) }
%%%%%%%%%%%%%%%%%%%%%%%%%%%%%%%%%%%
Its contribution reads\,\cite{NOVIKOV,ASNER}:
\beq
C_{4}\la  {\cal O}_{4}\ra\equiv (C_{40}+L_\mu C_{41})\la  \alpha_s G^2 \ra~~:~~ 
C_{40} = 2\pi a_s\ga 1 + \frac{175}{36} a_s\dr,~~C_{41} = \frac{9}{2}\pi a_s^2~.
\eeq
We shall use the value :
\beq
\gg=(6.35\pm 0.35)\times 10^{-2}\,{\rm GeV}^4,
\eeq
determined from light and heavy quark systems\,\cite{SNparam,SNparama,SNB8,SNREV15}. 
%%%%%%%%%%%%%%%%%%%%%%%%%%%%%%%%%%%%%%%%%%%%%%%%%%
\subsection*{\d The dimension-six ($n=3$) gluon condensate}
%%%%%%%%%%%%%%%%%%%%%%%%%%%%%%%%%%%%%%%%%%%%%%%%%%
Its  contribution reads to lowest order\,\cite{NOVIKOV} (see \cite{BAGAND6,ASNER} for the $\alpha_s$ correction for $n_f=0$\,\footnote{$C_{61}$ would be zero for $n_f=3$\,\cite{ZHANG}. We shall see in the analysis that the effect of this correction is negligible.}):
\beq
C_{6}\la  {\cal O}_{6}\ra= \ga C_{60}+L_\mu C_{61}\dr\ggg /Q^2~~:~~~~~~C_{60}=-a_s~,~~C_{61}=\frac{29}{4}a_s^2,
\eeq
with\,\cite{SNB8} :
\beq
\ggg=(8.2\pm 1.0)\,{\rm GeV}^2\gg~,
\eeq
which notably differs from the instanton liquid model estimate $\ggg\approx (1.5\pm 0.5)\,{\rm GeV}^2\gg$\,\cite{SVZ,NOVIKOV,NSVZ,SHURYAK} used in\,\cite{SNG,ASNER,ZHANG,FORKEL}.  However, the ratio of $\ggg$ over $\gg$ is in fair agreement with earlier lattice determination\,\cite{GIACOMO1}. 
%%%%%%%%%%%%%%%%%%%%%%%%%%%%%%%%%%%%%%%%%%%%%%%%%%
\subsection*{\d The dimension-eight ($n=4$) gluon condensate}
%%%%%%%%%%%%%%%%%%%%%%%%%%%%%%%%%%%%%%%%%%%%%%%%%%
Its contribution reads : 
\beq
C_{8}\la  {\cal O}_{8}\ra= C_{80}\gggg/Q^4~~ :~~~~~~C_{80}=4\pi \alpha_s
\eeq
with :
\beq
\gggg\equiv \Big{[}\la \ga \alpha_s f_{abc}G^a_{\mu\rho}G^b_{\nu\rho}\dr^2\ra +10\la \ga \alpha_s f_{abc}G^a_{\mu\nu}G^b_{\rho\lambda}\dr^2\ra\Big{]}\simeq 
%(0.55\pm 0.01)
k\ga\frac{15}{16}\dr \gg^2~:~~k=(1.5\pm 0.5)~,
\eeq
from  factorization\,\cite{NOVIKOV,NSVZ} where its validity has been questioned by  \,\cite{BAGAND8}.  Indeed, in the quark channel, the factorization hypothesis has been found to be largely violated for the four-quark condensates\,\cite{SNTAU,LNT,LAUNER}. To be conservative, we assume that the factorization is violated within the $k$-factor.
%%%%%%%%%%%%%%%%%%%%%%%%%%%%%%%%%%%%%%%%%%%%%%%
\subsection*{\b Beyond the standard SVZ-expansion}
%%%%%%%%%%%%%%%%%%%%%%%%%%%%%%%%%%%%%%%%%%%%%%%
\subsubsection*{\d The tachyonic gluon mass ($n=1$)}
%%%%%%%%%%%%%%%%%%%%%%%%%%%%%%%
To these standard contributions in the OPE, we can consider the one from a dimension-two tachyonic gluon mass contribution which is expected to phenomenologically parametrize the uncalculated  large order terms of the perturbative QCD series\,\cite{CNZa,CNZb,ZAKa,ZAKb} .
Its existence is supported by some AdS approaches\,\cite{ADS1,ADS2,ADS3}. 
This effect has been calculated explicitly in\,\cite{CNZa}\,:
\beq
C_2\la O_2\ra= -C_{21} L_\mu \lambda^2Q^2~ : ~~~~~~ C_{21}= {3}{a_s},
\eeq
where $\lambda^2$ is the tachyonic gluon mass determined from $e^+e^-\to$ hadrons data and the pion channel\,\,\cite{SND21,CNZa,TERAYEV}:
\beq
a_s\lambda^2\simeq -(6.0\pm 0.5)\times 10^{-2}~{\rm GeV}^2~.
\eeq
%%%%%%%%%%%%%%%%%%%%%
\subsubsection*{\d The direct instanton $(n\geq 5/2)$ } 
%%%%%%%%%%%%%%%%%%%%%%
In an instanton liquid model\,\cite{NSVZ,SHURYAK}, the direct instanton contribution is assumed to be dominated by the single instanton-anti-instanton contribution via a non-perturbative contribution to the perturbative Wilson coefficient\,\cite{FORKEL,ASNER}:
\beq
\psi_P(Q^2)\vert_{\bar I-I}=-32\,Q^4\int \rho^4 \Big{[} K_2\ga \rho\sqrt{Q^2}\dr\Big{]}^2 dn(\rho),
\eeq
which is opposite in sign with the one for the gluonium scalar correlator.
$K_2(x)$ is the modified Bessel function of the second kind. At this stage this classical field effect is beyond the SVZ expansion where the later assumes that one can separate without any ambiguity the perturbative Wilson coefficients from the non-perturbative condensate contributions. 

Besides the fact that it contributes in the OPE as $1/Q^5$, i.e.  it acts like other high-dimension condensates not taken into account in the OPE, the above instanton effect depends crucially on the (model-dependent) overall density $\bar n=\int_0^\infty d\rho\,n(\rho)$ and on its average size $\bar\rho=(1/\bar n)\int_0^\infty d\rho\,\rho\,n(\rho)$ which (unfortunately) are not quantitatively under a good control  ($\bar\rho$ ranges from 5 \,\cite{NSVZ} to 1.94,\cite{SNB8} and 1.65 GeV$^{-1}$\cite{SHURYAK}, while $\bar n \approx (0.5\sim 1.2)$\,fm$^{-4}$). They
contribute with a high power in $\rho$ to the spectral function ${\rm Im}\psi(t)$, which can be found explicitly in\,\cite{FORKEL,ASNER},  and behave as :
\bea
\frac{1}{\pi}{\rm Im}\psi_P (t)\vert_{\bar I-I}&& \buildrel t\to\infty\over \sim  -n\ga \rho\sqrt{t}\dr^{-5}\nnb\\
&& \buildrel t\to 0\over \sim -n\ga \rho\sqrt{t}\dr^{4}~.
\eea
However, it has been noticed in\,\cite{FORKEL} that this negative sign leads to a violation of positivity  and some inconsistencies for the sum rule analysis in this pseudoscalar channel.
As  such effects are quite inaccurate and model-dependent,  we shall not consider them explicitly  in our analysis.  Instead, 
%we shall test (a posteriori) its effect  by comparing 
an eventual deviation of our results within the standard SVZ expansion 
from some experimental data or/and or some other alternative estimates (Low-Energy Theorems (LET), Lattice calculations,...) may signal the need of such (beyond the standard OPE) effects in the analysis.   

One should mention that the approach within the standard SVZ OPE and without a direct instanton effect used in the :

\hspace*{0.5cm} --  $U(1)_A$ channel has predicted successfully the value of the topological charge, its slope and the $\eta'$-mass and decay constant\,\cite{SNU1,SNU2,SNG0,SNG1,SHORE}. 

\hspace*{0.5cm} -- Pseudoscalar pion and kaon channels have reproduced successfully the value of the light quark masses where we have also explicitly shown\,\cite{SNL} that  the direct instanton effect induces a relatively small correction  contrary to some  vigorous claims  in the literature (see e.g.\,\cite{IOFFE}). 
%%%%%%%%%%%%%%%%%%%%%%%%%%%%%%%%%%%%%%%%%%%
%{\scriptsize
\begin{table}[hbt]
\setlength{\tabcolsep}{2.8pc}
 \caption{{\small QCD input parameters from recent QSSR analysis based on stability criteria. $k$ measures the violation of factorization. }} 
 \vspace*{-0.25cm}
%\tbl{
%}
{\small
 % \begin{tabular}{lll}
 \begin{center}
 {\begin{tabular}{@{}llll@{}}
&\\
%\hline
\hline
%\\
Parameter&Value& Ref.    \\
%\\
\hline
%\\
$\alpha_s(M_Z)$ & $0.1181(16)(3)\Lrar \Lambda=(340\pm 28)$ MeV & \cite{SNparam,SNparama}\\
$\lambda^2\times 10^2$ [GeV$^2$]& $-6.0(5)$ & \cite{SND21,TERAYEV}\\
$\la\alpha_s G^2\ra\times 10^2$ [GeV$^4$]& $6.35(35)$ & \cite{SNB8,SNREV15}\\
$\la g^3  G^3\ra / \la\alpha_s G^2\ra$ [GeV$^2$]& $8.2(1.0)$ &\cite{SNB8,SNREV15,GIACOMO1}\\
$\la \alpha_s^2 G^4\ra$&$k\frac{15}{16}\la \alpha_sG^2\ra^2~~~: ~~~k=(1.5\pm 0.5)$&\cite{NSVZ}\\
%\hline
\hline
\end{tabular}}
\end{center}
}
\label{tab:param}
\end{table}
\vspace*{-0.55cm}
%%%%%%%%%%%%%%%%%%%%%%%%%%%%%%%%%%%%%%%%%%%

%%%%%%%%%%%%%%%%%%%%%%%%%%%%%%%%%%%%%%%%%%%
\section{The Inverse Laplace transform sum rules}
%%%%%%%%%%%%%%%%%%%%%%%%%%%%%%%%%%%%%%%%%%%
From its QCD asymptotic behaviour $\sim (-q^2)^2{\rm Log}(-q^2/\mu^2)$ one can write a twice subtracted dispersion relation:
\beq
\psi_P(q^2)=\psi_P(0) +q^2\psi'_P(0)+\frac{q^4}{\pi} \int_0^\infty \frac{dt}{t^2}\,\frac{{\rm Im}\psi_P(t)} {(t-q^2-i\epsilon)}~.
\eeq

\b Following standard QSSR techniques\,\cite{SVZ,SNB1,SNB2}, one can derive from it different form of the sum rules.  In this paper, we shall work with the Exponential or Borel \,\cite{SVZ,BELLa,BERTa} or Inverse Laplace transform\,\cite{SNR} Finite Energy sum rule(LSR)\footnote{The name Inverse Laplace transform has been attributed due to the fact that perturbative radiative corrections have this property.}\,:
\beq
{\cal L}^c_n(\tau)=\int_0^{t_c} \hspace*{-0.25cm}dt\,t^n\, e^{-t\tau}\frac{1}{\pi} {\rm Im}\psi_P(t)~: n=-2,-1,0,1,...,4,
\eeq
and the corresponding ratios of sum rules :
\beq
{\cal R}^c_{n+l~n}(\tau)\equiv \frac{{\cal L}^c_{n+l}(\tau)}{{\cal L}^c_n(\tau)}~,
\eeq
where $\tau$ is the Laplace sum rule variable. In the duality ansatz :
\beq
\frac{1}{\pi} {\rm Im}\psi_P(t)=2\sum_P f_P^2M_P^4\,\delta(t-M_G^2)+\theta (t-t_c) {\rm "QCD\, continuum"}~: \la 0\vert (8\pi) Q(x)\vert P\ra=\sqrt{2}f_P M_P^2, 
\label{eq:spectral}
\eeq
where the $f_P$ is the  resonance coupling normalized as $f_\pi=93$ MeV;  $M_P$ is the mass while the "QCD continuum" comes from the discontinuity $ {\rm Im}\psi_P(t)\vert_{QCD}$ of the QCD diagrams from the continuum threshold $t_c$.  In the {\it ``One narrow resonance $\oplus$ QCD continuum"} parametrization of the spectral function\,\footnote{Finite width effect has been shown in the example of the large width $\sigma$ meson to give a negligble correction\,\cite{VENEZIA,SNG} and will not be considered in this paper.}:
\beq
{\cal R}^c_{n+l~n}(\tau)\simeq M_P^2~.
\eeq

\b ${\cal L}_{(-2,-1)}$ have been used to estimate the topological charge, its slope and the $\eta'$ parameter within the minimal duality ansatz which will be reviewed and dscuused in Section\,\ref{sec:lm2}. 

\b To get ${\cal L}_0$, we take the Inverse Laplace transform of the 1st superconvergent 3rd derivative of the two-point correlator. In this way, we obtain\,:
\beq
{\cal L}^c_{0}(\tau)=\tau^{-2}\,2\hspace*{-0.25cm}\sum_{n=0,2,\cdots}\hspace*{-0.25cm}D^{0}_n ~,
\eeq
with [$L_\tau\equiv {\rm Log}\,{(\tau\mu^2)}$]\,: 
\bea
D^{0}_0&=&\Big{[}2C_{00}-2C_{01}(3-2\gamma_E-2L_\tau)-6C_{02}\big{[}1-3\gamma_E+\gamma^2-\pi^2/6+(-3+2\gamma_E)L_\tau+L_\tau^2\big{]}\Big{]}(1-\rho_2),\nnb\\
D^{0}_2&=&-\frac{C_{21}}{2}\lambda^2\tau(1-\rho_1),\nnb\\
D^{0}_4&=&-{C_{41}}\gg\tau^2,\nnb\\
%D^{0}_4&=&-\Big{[}C_{40}-\frac{C_{41}}{36}\big{[} 55+6(\gamma_E+L)\big{]}\Big{]}\gg\tau^3,\nnb\\
D^{0}_6&=&\big{[}C_{60}-C_{61}(\gamma_E-L_\tau)\big{]}\ggg\tau^3,\nnb\\
D^{0}_8&=&C_{80}\ga\frac{15}{16}\dr k\,\gg^2\tau^4~.
\label{eq:l0}
\eea

\b The other higher degrees sum rules ${\cal L}^c_n(\tau)$ for $n\geq 1$ can be deduced from the $n^{th}$ $\tau$-derivative of ${\cal L}^c_0(\tau)$:
\beq
{\cal L}^c_n(\tau)=(-1)^n\frac{d^n}{d\tau^n} {\cal L}^c_0(\tau)~.
\eeq

\b These superconvergent sum rules obey the homogeneous renormalization group equation (RGE):
\beq
\Big{\{}-\frac{\partial }{\partial t} +\beta(\alpha_s)\alpha_s \frac{\partial }{\partial \alpha_s}\Big{\}}{\cal L}^c_n(e^t \tau,\alpha_s)=0~,
\eeq
where $t\equiv (1/2)L_\tau$. The renormalization group improved (RGI) solution is:
\beq
{\cal L}^c_n(e^t \tau,\alpha_s)= {\cal L}^c_n(t=0,\bar\alpha_s(\tau))~,
\eeq
where $\bar\alpha_s(\tau)$ is the QCD running coupling. 

\b In the following analysis, we shall work with the family of sum rules having degrees less or equal to 4. In so doing, we shall select the sum rules which present stability (minimum or inflexion point) in the sum rule variable $\tau$ and in the continuum threshold $t_c$ such that we can extract optimal information from the analysis. 
 %%%%%%%%%%%%%%%%%%%%%%%%%%%%%%%%%%%%%
\section{The $\eta'$ mass and decay constant \label{sec:lm2}}
 %%%%%%%%%%%%%%%%%%%%%%%%%%%%%%%%%%%%%
 %%%%%%%%%%%%%%%%%%%%%%%%%%%%%%%%%%%%%%%
\begin{figure}[hbt]
%\vspace*{-0.25cm}
\begin{center}
\vspace{-0.25cm}
\includegraphics[width=8.cm]{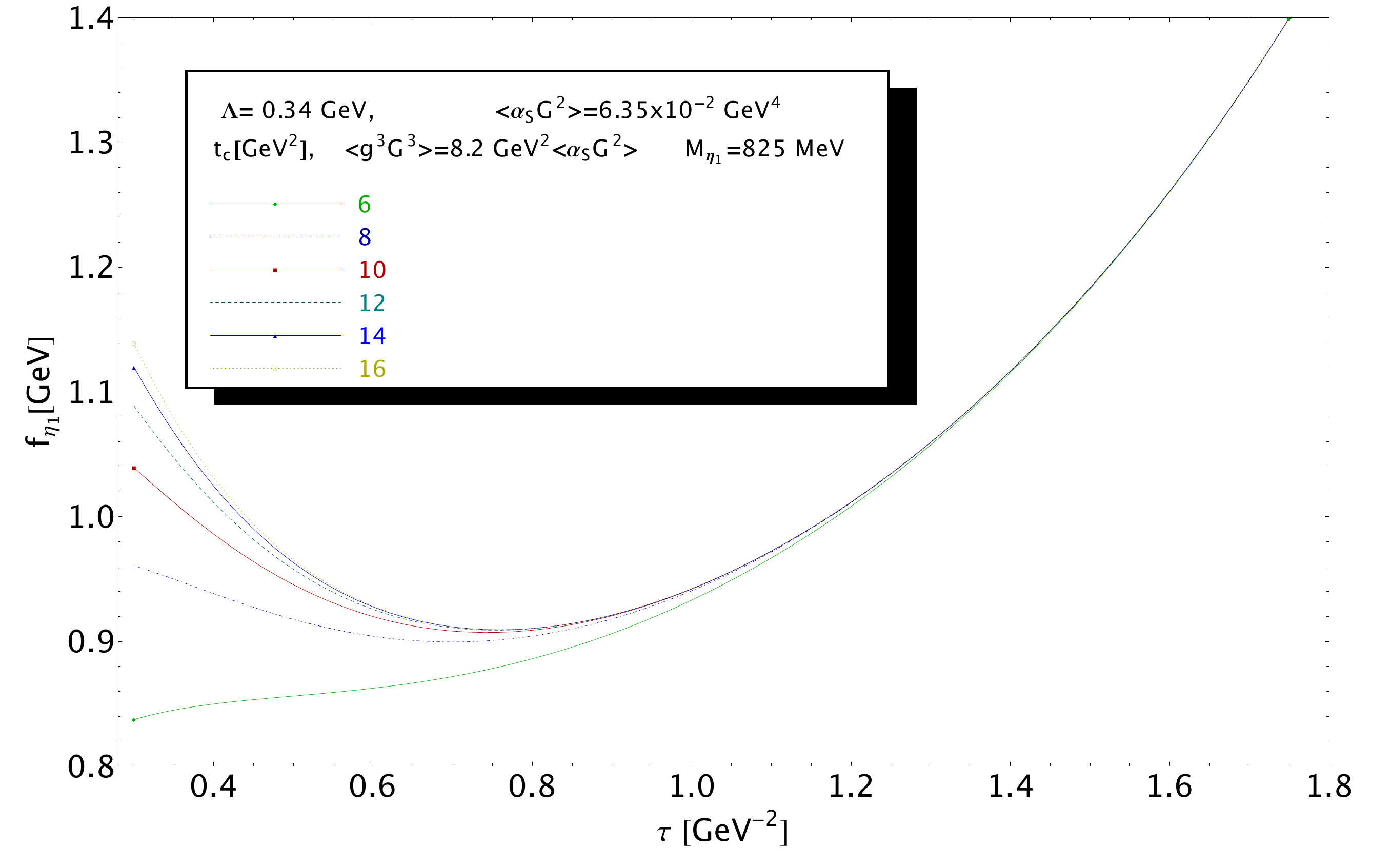}
\vspace*{-0.5cm}
\caption{\footnotesize  The decay constant $f_{\eta_1}$ from ${\cal L}_{-1}^c$  as a function of $\tau$ at N2LO using $M_{\eta_1}=825$ MeV  and the QCD inputs  in Table\,\ref{tab:param}.} 
\label{fig:feta1}
\end{center}
%\vspace*{-0.75cm}
\end{figure} 
%%%%%%%%%%%%%%%%%%%%%%%%%%%%%%%%%%%%%%%

 Keeping only the $\eta'$ and/or the lowest mass gluonium contributions in the parametrization of  the spectral function (Eq.\,\ref{eq:spectral}),  ${\cal L}^c_{-1}$ and ${\cal L}^c_{0}$  have been used earlier to look for the gluon component of the $\eta'$ mass and coupling and to estimate the topological charge  $\chi(0)$ and its slope\,$\chi'(0)$\,\cite{SNU1,SNU2,SNG0,SNG1,SHORE,SHORE1}. The LSR results in pure Yang-Mills  fairly agree with the ones from large $N_c$\,\cite{WITTEN,VENEZIANO,DIVECCHIA} and lattice calculations\,\cite{GIACOMO}. %The QCD expressions of these two sum rules can be deduced from previous quoted papers and the N2LO one from the scalar analogue in Ref.\,\cite{SNS21}. 
 %As, in this case, the sum rule depends also on the  the topological susceptibility $\chi(0)$, there is an iterative loop procedure for extracting these parameters. Therefore, 
Here, we use the sum rule  ${\cal L}^c_{-1}$ in the chiral limit where  $\chi(0)\equiv\psi_P(0)/(8\pi)=0$:
  \beq
  {\cal L}^c_{-1}\equiv\int_0^{t_c} \frac{dt}{t}\,e^{-t\tau} \frac{1}{\pi}\,{\rm Im}\psi_P(t) =  {\cal L}^c_{-1}\vert_{\rm QCD}+ \psi_P(0)~,
  \label{eq:lm1}
  \eeq
  where :
  \beq
{\cal L}^c_{-1}\vert_{QCD}=\tau^{-2}\,2\hspace*{-0.25cm}\sum_{n=0,2,\cdots}\hspace*{-0.25cm}D^{-2}_n~,
\eeq
with:
\bea
D^{-1}_0&=&\Big{[}C_{00}+2C_{01}(1-\gamma_E-L_\tau)+3C_{02}\big{[}(\gamma_E+L_\tau)^2-2(\gamma_E+L_\tau)-\frac{\pi^2}{6}\big{]}\Big{]}(1-\rho_1)\nnb\\
D^{-1}_2&=&\tau\,C_{21}\lambda^2(1-\rho_0)\,,\nnb\\
D^{-1}_4&=&-\tau^2\Big{[}C_{40}-C_{41}(\gamma_E+L_\tau)\Big{]}\gg,\nnb\\
%\eea
%\end{document}
D^{-1}_6&=&-{\tau^3}\Big{[}C_{60}+C_{61}\ga 1-\gamma_E-L_\tau\dr\Big{]}\ggg,\nnb\\
D^{-1}_8&=&-\frac{\tau^4}{2}C_{80}\gg^2~,
\label{eq:lm1b}
\eea
for extracting the decay constant $f_{\eta_1}$. We have attempted to extract the mass $M_{\eta_1}$ using the low degree ratio of moments ${\cal R}_{0-1}$ but fail due to the absence of $\tau$ stability. Then, we use the GMO mass formula quoted in\,\cite{SHORE} derived  in the massless pion limit and assuming a SU3 symmetry for the decay constants from\,\cite{VENEZIANO}\,:
  \beq
 M_{\eta_1}^2\simeq M_{\eta'}^2-\frac{2}{3}M_K^2=(870\,{\rm MeV})^2.
 \eeq
  An alternative derivation of $M_{\eta_1}$ in pure Yang-Mills from $1/N_c$ expansion by\,\cite{WITTEN,VENEZIANO,DIVECCHIA}  and corrected in\,\cite{VENEZIA} by including 3 massless quarks loops contribution through the $\beta$ function [${\cal O}(n_f/N_c)$ correction] leads to\,:
  \beq
 M_{\eta_1}^{ n_f=3}\simeq\ga\frac{\beta_1^{ n_f=3}}{\beta_1^{ n_f=0}}\dr^{1/2}\hspace*{-0.5cm}\times
 M_{\eta_1}^{YM}\simeq 779\,{\rm MeV}~~~: ~~ ~\beta_1^{n_f}=-\frac{1}{2}\ga 11-\frac{2}{3}n_f\dr,
 \eeq
 with :
 \beq
 M_{\eta_1}^{YM}\simeq \frac{1}{f_\pi}\ga 6\frac{\partial}{\partial\theta}\la Q(x)\ra_{\theta=0}  \equiv 
 -6\chi(0)\vert_{YM} \dr^{1/2}
 :~~~\chi(0)\vert_{YM}\simeq-\ga 180\,\rm{MeV}\dr^4~~{\rm and}~~~f_\pi=92.2~{\rm MeV}. 
 \eeq
 In the following analysis, we shall use the mean of the two determinations\,:
  \beq
 M_{\eta_1}\simeq 825(45)~{\rm MeV},
 \label{eq:meta1}
 \eeq
 with the (conservative) error from its distance to the two former. 
The analysis for N2LO is shown in Fig.\,\ref{fig:feta1} where, at the $\tau$-minimum$\,:\tau=(0.68;0.74)\pm 0.04 \,\rm GeV^{-2}$ corresponding to $t_c=8$ (beginning of $\tau$-stability)  to 14 GeV$^2$ ($t_c$-stability), one deduces the estimate:
 \beq
  f_{\eta_1}
  %&=&711(19)_{t_c}(2)_\tau(36)_{\Lambda}(71)_{\lambda^2}(2)_{G^2}(2)_{G^3}(0)_{G^4}= 711(82)~\rm MeV~: ~~~~NLO\nnb\\
 =905(5)_{t_c}(1)_\tau(43)_{\Lambda}(53)_{\lambda^2}(3)_{G^2}(2)_{G^3}(0)_{G^4}(24)_{M_{\eta_1}}= 905(72)~\rm MeV.
\label{eq:feta}
 \eeq
 where one should note that $f_{\eta_1}$ here is not exactly the one from e.g. $J/\psi\to \gamma\eta',~\eta'\to \gamma\gamma$\,\cite{SHORE2}.  One can notice that\,:
 %,\footnote{I wish to thank Graham Shore and Gabriele Veneziano for some discussions on the results and on their effects on  the so-called proton spin problem in the following Section\,\ref{sec:spin}.}\,:
 
 \d The N2LO correction is large and has increased the value of $f_{\eta_1}$ from NLO by about a factor 2. The large N2LO effects to $f_{\eta_1}$ requires a further control of the size of the higher order PT corrections. We have considered in\,\cite{CNZa} that the tachyonic gluon mass gives an estimate of such uncalculated higher order terms of PT or it is equivalent to say that the  PT series grows geometrically\,\cite{CNZb}. One can check that the tachyonic gluuon mass  decreases $f_{\eta_1}$ by 53 MeV which we consider (here and in the following) as an estimate of the errors due to the truncation of the PT series but we do not include it in the central value of $f_{\eta_1}$.  Considering this error estimate at its face value, one may consider that the PT series reach its asympotic value at N2LO. 

\d The value of $M_{\eta_1}$ and of $f_{\eta_1}$ in Eqs.\,\ref{eq:meta1} and \,\ref{eq:feta} will be used as inputs in the rest of the paper.

 %%%%%%%%%%%%%%%%%%%%%%%%%%%%%%%%%%%%%
\section{The slope $\chi'(0)$ of the topological charge and the proton spin  in the chiral limit} \label{sec:spin}
 %%%%%%%%%%%%%%%%%%%%%%%%%%%%%%%%%%%%%
  %%%%%%%%%%%%%%%%%%%%%%%%%%%%%%%%%%%%%%%
\begin{figure}[hbt]
%\vspace*{-0.25cm}
\begin{center}
\vspace{-0.25cm}
\includegraphics[width=8.cm]{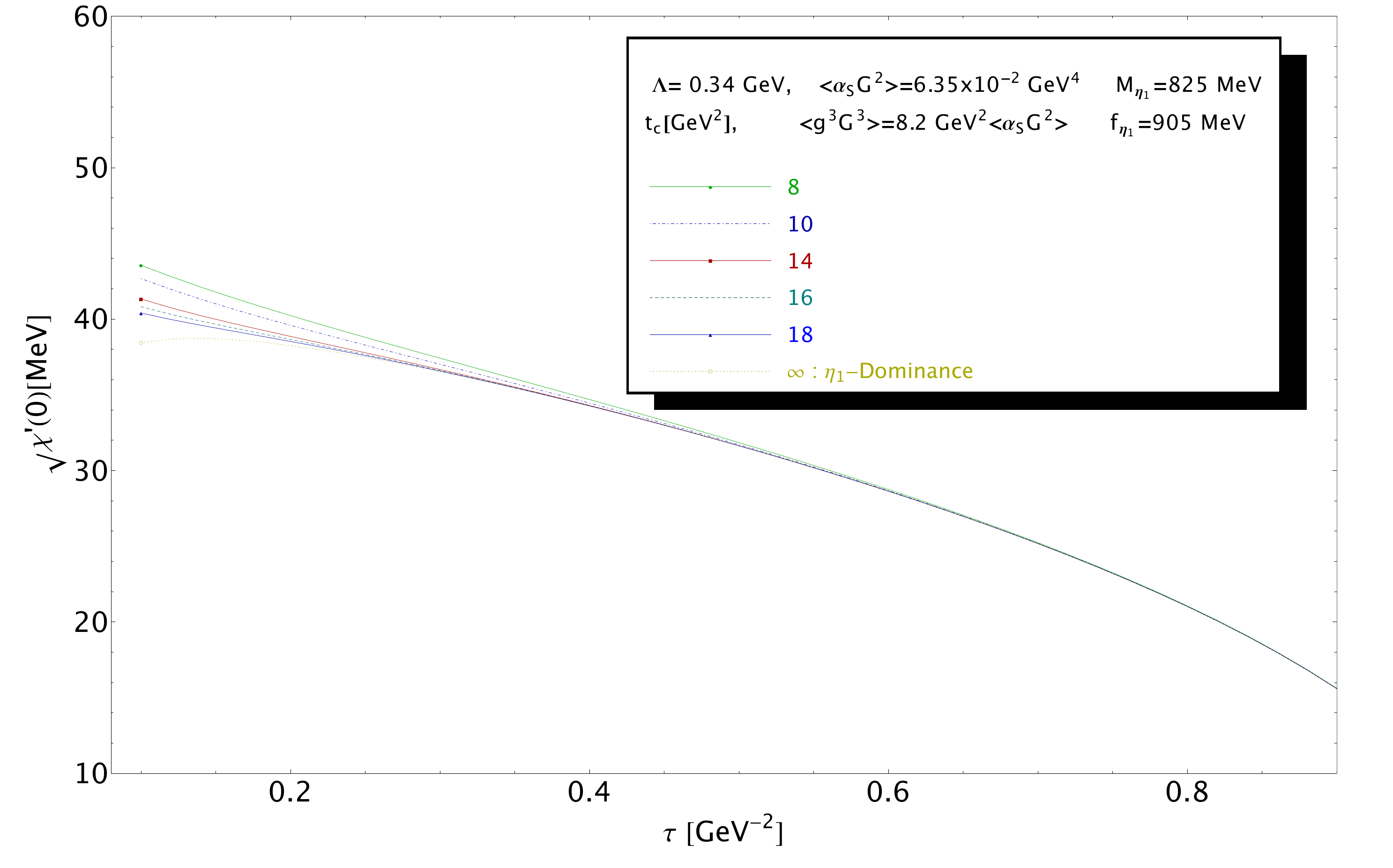}
\vspace*{-0.5cm}
\caption{\footnotesize  The slope of the topological charge $\chi'(o)$ from ${\cal L}_{-2}^c$  as a function of $\tau$ at N2LO using $f_{\eta_1}=881$ MeV.} 
\label{fig:chiprim-N2LO}
\end{center}
%\vspace*{-0.75cm}
\end{figure} 
%%%%%%%%%%%%%%%%%%%%%%%%%%%%%%%%%%%%%%%
  %%%%%%%%%%%%%%%%%%%%%%%%%%%%%%%%%%%%%%%
\begin{figure}[hbt]
%\vspace*{-0.25cm}
\begin{center}
\vspace{-0.25cm}
\includegraphics[width=8.cm]{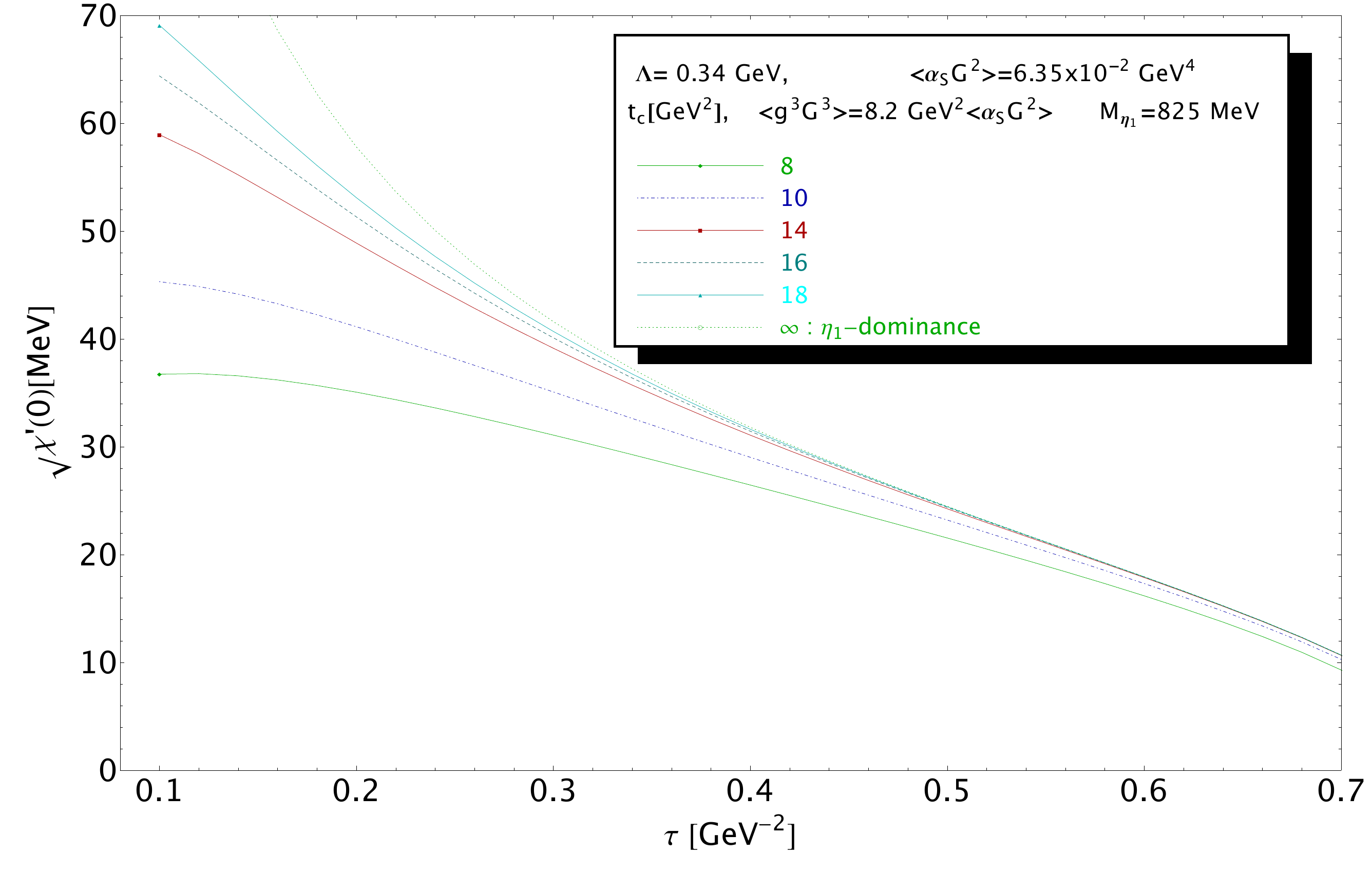}
\vspace*{-0.5cm}
\caption{\footnotesize  The slope of the topological charge $\chi'(o)$ from ${\cal L}_{-2}^c$  as a function of $\tau$ at N2LO using the QCD expression of $f_{\eta_1}$ given in Eqs.\,\ref{eq:lm1} to \,\ref{eq:lm1b}.}
\label{fig:chiprim-N2LO2}
\end{center}
\vspace*{-0.75cm}
\end{figure} 
%%%%%%%%%%%%%%%%%%%%%%%%%%%%%%%%%%%%%%%

%%%%%%%%%%%%%%%%%%%%%%%%%%%%%%%%%%%%%%%
\subsection*{\b Sum rule estimate of $\chi'(0)$ at N2LO  in the chiral limit}
%%%%%%%%%%%%%%%%%%%%%%%%%%%%%%%%%%%%%%%
$\chi'(0)$ has been estimated in pure Yang-Mills to be\,\cite{SNG1}:
\beq
\chi'(0)\vert_{YM}=-[(7\pm 3)\,\rm MeV]^2
\eeq
in agreement with the lattice result $-[(9.8\pm0.9)\,\rm MeV]^2$\,\cite{GIACOMO}. In the following, we shall updated its NLO estimate in the chiral limit\,\cite{SHORE} by including N2LO PT corrections and NLO ones to the condensate contributions.
 We shall use the previous value of  $M_{\eta_1}$ and the value or the QCD expression of $f_{\eta_1}$ into the twice subtracted sum rule:
 \beq
  \psi'_P(0)\vert_{\rm LSR} =\int_0^{t_c} \frac{dt}{t^2}\,e^{-t\tau} \frac{1}{\pi}\,{\rm Im}\psi_P(t)-  {\cal L}^c_{-2}\vert_{QCD}+
 \tau\, \psi_P(0)~,
  \eeq
 % \vspace*{-0.25cm}
  where :
  \beq
{\cal L}^c_{-2}\vert_{QCD}=\tau^{-1}\,2\hspace*{-0.25cm}\sum_{n=0,2,\cdots}\hspace*{-0.25cm}D^{-2}_n~,
\eeq
with:
\bea
D^{-2}_0&=&\Big{[}C_{00}-2C_{01}(\gamma_E+L_\tau)+3C_{02}\big{[}(\gamma_E+L_\tau)^2-\frac{\pi^2}{6}\big{]}\Big{]}(1-\rho_0)\nnb\\
D^{-2}_2&=&-\tau\,C_{21}\gamma_E\lambda^2\, ,\nnb\\
D^{-2}_4&=&\tau^2\Big{[}C_{40}+C_{41}(1 -\gamma_E)\Big{]}\gg,\nnb\\
%\eea
%\end{document}
D^{-2}_6&=&\frac{\tau^3}{2}\Big{[}C_{60}+C_{61}\ga\frac{3}{2}-\gamma_E-L_\tau\dr\Big{]}\ggg,\nnb\\
D^{-2}_8&=&\frac{\tau^4}{6}C_{80}\gg^2~,
\eea
%where $\psi_P(0)=0$ in the chiral limit. 

 We show the N2LO analysis  from ${\cal L}_{-2}^c$ where, in Fig.\,\ref{fig:chiprim-N2LO}, we use the value of $f_{\eta_1}$ in Eq.\,\ref{eq:feta} and in Fig.\,\ref{fig:chiprim-N2LO2} its QCD expression from Eq.\,\ref{eq:lm1} where one should note  that the sum of higher states contributions to the spectral function is (automatically) included in this second option.  One can notice a better $\tau$-stability (clearer inflexion point) for the 2nd case at $\tau \simeq (0.50\pm 0.04)\,\rm GeV^{-2}$ and for  $t_c\simeq (14\sim 18)\,\rm GeV^2$, at which, we deduce:  
 \beq
\rm \sqrt{\chi'(0)}\vert_{N2LO} \equiv  \frac{1}{8\pi} \sqrt{\psi'_P(0)}=  24.3(2.8)_\tau (0.2)_{t_c}(1.6)_{\Lambda}(0.8)_{\lambda^2}(0.2)_{G^2}(0.4)_{G^3}(0.1)_{G^4}(0.4)_{M_{\eta_1}}=24.3(3.4)~\rm MeV.
\label{eq:chiprim}
  \eeq 
  The NLO analysis gives curves very similar to the N2LO ones. At the corresponding inflexion point $\tau\simeq (0.44\pm 0.04)\, \rm GeV^{-2}$, one obtains:
  \beq
  \rm \sqrt{\chi'(0)}\vert_{NLO} = 18.2(2.5)_\tau (0.2)_{t_c}(1.1)_{\Lambda}(0.9)_{\lambda^2}(0.4)_{G^2}(0.6)_{G^3}(0.1)_{G^4}(0.4)_{M_{\eta_1}}=18.2(3.0)~\rm MeV.
  \eeq
 These new values agree within the errors  with the NLO one $22.3(4.8)$ MeV obtained from Laplace sum rule (LSR) in\,\cite{SHORE} using different values of the QCD input parameters and a slightly different QCD expression due to some errors in the $\alpha_s$ PT corrections  giving by\,\cite{KATAEV} explaining that the $\chi'(0)$ curves present minimum there.  
 %%%%%%%%%%%%%%%%%%%%%%%%%%%%%
 \subsection*{\b Effect on the proton spin}
 %%%%%%%%%%%%%%%%%%%%%%%%%%%%%
 It has been shown by\,\cite{SHORE2,SHORE1} (see also\,\cite{TARASOV}) that the singlet form factor $G_A^{(0)}(Q^2)$ appearing in the first moment of the polarised proton structure function $g_1^P$ (Ellis-Jaffe sum rule\,\cite{JAFFE}) can be related in the chiral limit ($m_q=0$) to the slope of the topological charge $\chi'(0)$ as:
 \beq
G_A^{(0)}(Q^2)\bar u\gamma_5 u = \frac{1}{2M_p} (2n_f)\sqrt{\chi'(0)}\,\Gamma_{\Phi_{5R}\bar PP},
\eeq
 where $\Phi_{5R}$ is the renormalized bilinear quark current and $\Gamma_{\Phi_{5R}\bar PP}$ the renormalization group (RG) invariant scale independent proper vertex. In the chiral limit and using the OZI approximation, one would obtain\,\cite{SHORE}:
 \beq
 \Gamma_{\Phi_{5R}\bar PP}\vert_{\rm OZI}=\sqrt{2} g_{\eta_8 PP}\,(\bar u\gamma_5 u), ~~~~~~~~~ \sqrt{\chi'(0)}\vert_{\rm OZI}={f_\pi}/{\sqrt{6}}=38~{\rm MeV},
 ~~~~~~~~~  G_A^{(0)}\vert_{\rm OZI}=0.579\pm 0.021. 
  \eeq
  while from our analysis:
  \beq
  G_A^{(0)}\vert_{\rm LSR}(Q^2=10~\rm GeV^2)= G_A^{(0)}\vert_{\rm OZI} 
  \frac{\sqrt{{\chi'(0)}}\vert_{\rm LSR}}
 {\sqrt{{\chi'(0)}}\vert_{\rm OZI}}= (0.340\pm 0.050),
%  \equiv \Delta\Sigma=\Delta u + \Delta d + \Delta s=3F-D=0.579\pm 0.021~,
\label{eq:ga}
\eeq
after running $\chi'(0)\vert_{\rm LSR}$ from 2 to 10 GeV$^2$  implying:  
\beq
\sqrt{{\chi'(0)}}\vert_{\rm LSR}(Q^2=10\,\rm GeV^2)=(22.5\pm 3.1)~{\rm MeV}.
\eeq
Using the previous value of  $G_A^{(0)}\vert_{\rm LSR}$ and:
$
G_A^{(3)} = 0.625\pm 0.004,~~G_A^{(8)} = (0.167\pm 0.006),
$
from hyperon and $\beta$-decays\,\cite{HYPERON}, 
the first moment of the polarised proton structure function gives: 
\beq
\Gamma^p_1(Q^2=10 ~\rm GeV^2)\equiv \int_0^1 dx\,g_1^P(x,Q^2)= (0.144\pm 0.005),
\label{eq:spin1}
\eeq
in excellent agreement with the earlier data\,\cite{EMC} and experimental world average\,\cite{SMC2} :
\beq
\Gamma^p_1(Q^2=10 ~\rm GeV^2)\vert_{\rm exp}=0.145\pm 0.014~~~~\Lrar ~~~~ G_A^{(0)}\vert_{\rm exp} = 0.35\pm 0.12,
\eeq
and more recent data from COMPASS\,\cite{COMPASS} and HERMES\,\cite{HERMES}\,:
\beq
G_A^{(0)}\vert_{\rm exp} = 0.33\pm 0.04,
\eeq
which are lower than the OZI value by a factor 1.76. 

%%%%%%%%%%%%%%%%%%%%%%%%%%%%%%%%%%%%%%%%%%
%\section*{Acknowledgements}
%%%%%%%%%%%%%%%%%%%%%%%%%%%%%%%%%%%%%%%%%
%It is a pleasure to thank Garaham Shore and Gabriel Veneziano for several exchanges which have lead to the improvement and to the extension of the discussions on $\chi'(0)$ and the proton spin in Sections\,\ref{sec:lm2} and \,\ref{sec:spin}. 

 %%%%%%%%%%%%%%%%%%%%%%%%%%%%%%%
% \vfill\eject
% \input{bib_pseudo.tex}
% \end{document}
 %%%%%%%%%%%%%%%%%%%%%%%%%%%%%%%%%%%%%%%%%%%
 %%%%%%%%%%%%%%%%%%%%%%%%%%%%%%%
% \vfill\eject
% \input{bib_pseudo.tex}
 %\end{document}
 %%%%%%%%%%%%%%%%%%%%%%%%%%%%%%%%%%%%%%%%%%%
 %%%%%%%%%%%%%%%%%%%%%%%%%%%%%%%%%%%%%
\section{On some previous gluonium mass from the ratio of moments ${\cal R}_{10}$\label{sec:r10}}
 %%%%%%%%%%%%%%%%%%%%%%%%%%%%%%%%%%%%%
\b The extraction of the lowest gluonium mass from the sum rule within the minimal duality ansatz ``{\it One resonance $\oplus$ QCD continuum}" parametrization of the spectral function often comes from the the ratio of moments ${\cal R}_{10}$\,\cite{SNG,SNG0,ASNER} with the most recent result to N2LO within the standard SVZ-expansion from Laplace sum rule\,\cite{ASNER} and without including the $\eta'$ contribution:
\beq
M_{P_2}\vert_{\rm N2LO}=  (2.3\pm 0.2)~{\rm GeV} ~~~~~~f_{P_2}=(0.21\pm 0.04)~{\rm GeV}~~~~~{\rm for}~~~~\Lambda=(0.15\pm 0.05)~{\rm GeV}~.
\label{eq:15}
\eeq
Similar results using Laplace sum rules have been obtained in\,\cite{FORKEL} by adding the contribution of instanton-anti-instanton in the OPE and by including the contribution of the $\eta'$: 
\beq
M_{P_2}\vert_{\rm N2LO}=  (2.2\pm 0.2)~{\rm GeV},~~~~~~f_{P_2}=(0.42\pm 0.18)~{\rm GeV} ~~~~{\rm for}~~~~\Lambda=0.20~{\rm GeV}~.
\label{eq:20}
\eeq
The Gaussian sum rule including instanton-anti-instanton but without the $\eta'$ leads to\,\cite{ZHANG} :
\beq
M_{P_2}\vert_{\rm N2LO}=  (2.65\pm 0.33)~{\rm GeV},~~~~~~\Gamma_{P_2}\leq 0.54~{\rm GeV} ~~~~{\rm for}~~~~\Lambda=0.3~{\rm GeV}~,
\label{eq:30}
\eeq
The previous N2LO results can be compared with the LO one (independent of the $\Lambda$ value) \,\cite{SNG0}:
\beq
M_{P_2}\vert_{\rm LO}=(1.70\pm 0.06)~{\rm GeV}~,
\eeq
and to the NLO one {\rm for} $\Lambda=(0.375\pm 0.125)$~{\rm GeV}\,\cite{SNG}:
\beq
M_{P_2}\vert_{\rm NLO}=(2.05\pm 0.19)~{\rm GeV}.
\eeq
 %%%%%%%%%%%%%%%%%%%%%%%%%%%%%%%%%%%%%%%
\begin{figure}[hbt]
%\vspace*{-0.25cm}
\begin{center}
%\centerline {\hspace*{-6.cm} \bf a)\hspace*{10cm} }
\vspace{-0.25cm}
%\includegraphics[width=10cm]{L-1_OPE.pdf} \\
%\centerline {\hspace*{-6.cm} \bf b)\hspace*{10cm}}
\includegraphics[width=8.cm]{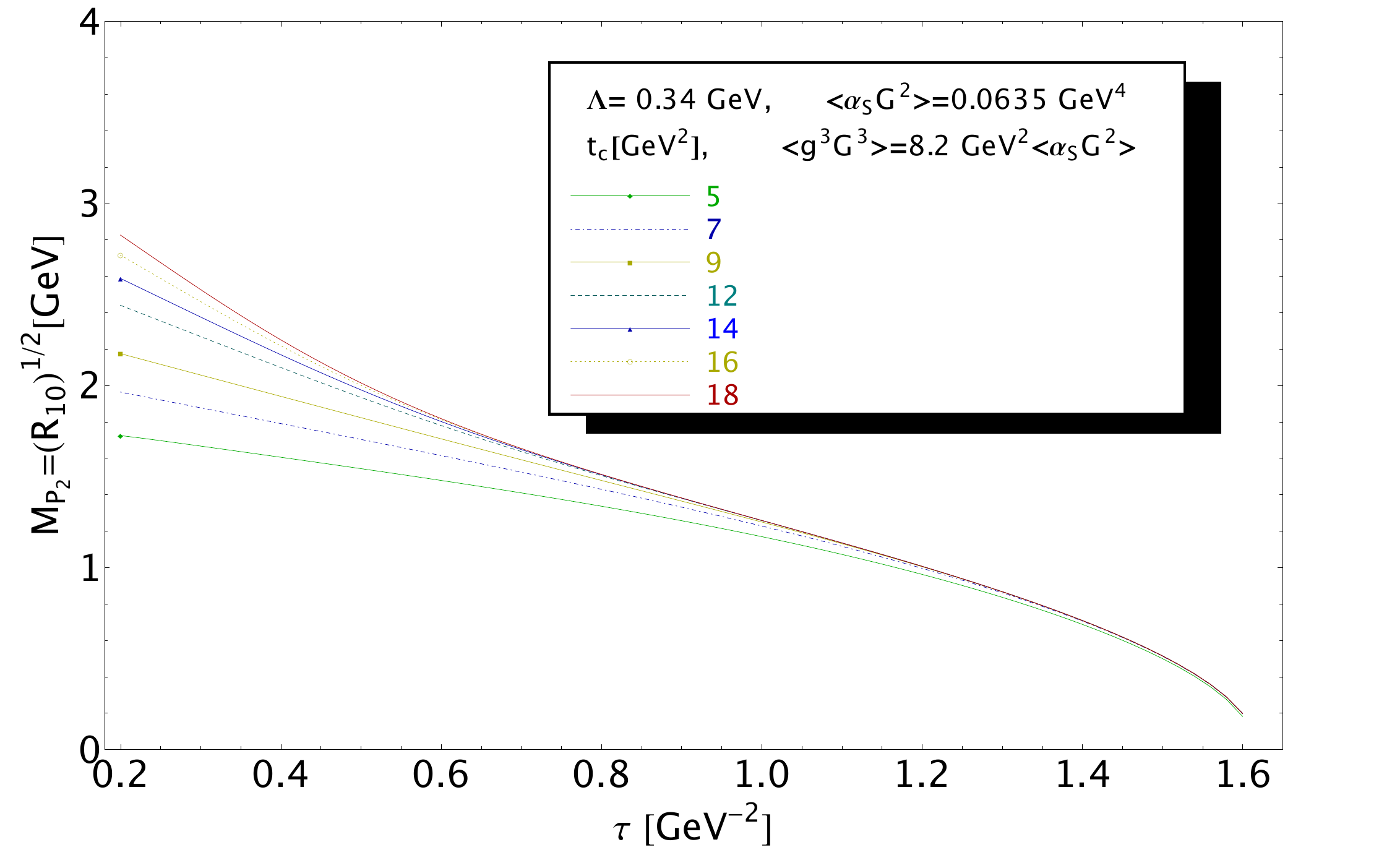}
\vspace*{-0.5cm}
\caption{\footnotesize  $M_{P_2}$ from ${\cal R}^c_{10}$  as a function of $\tau$ at N2LO for different values of $t_c$ and for $\Lambda=0.34$ GeV.} 
\label{fig:10}
\end{center}
%\vspace*{-0.75cm}
\end{figure} 
%%%%%%%%%%%%%%%%%%%%%%%%%%%%%%%%%%%%%%%

 %%%%%%%%%%%%%%%%%%%%%%%%%%%%%%%%%%%%%%%
\begin{figure}[hbt]
%\vspace*{-0.25cm}
\begin{center}
\includegraphics[width=8.cm]{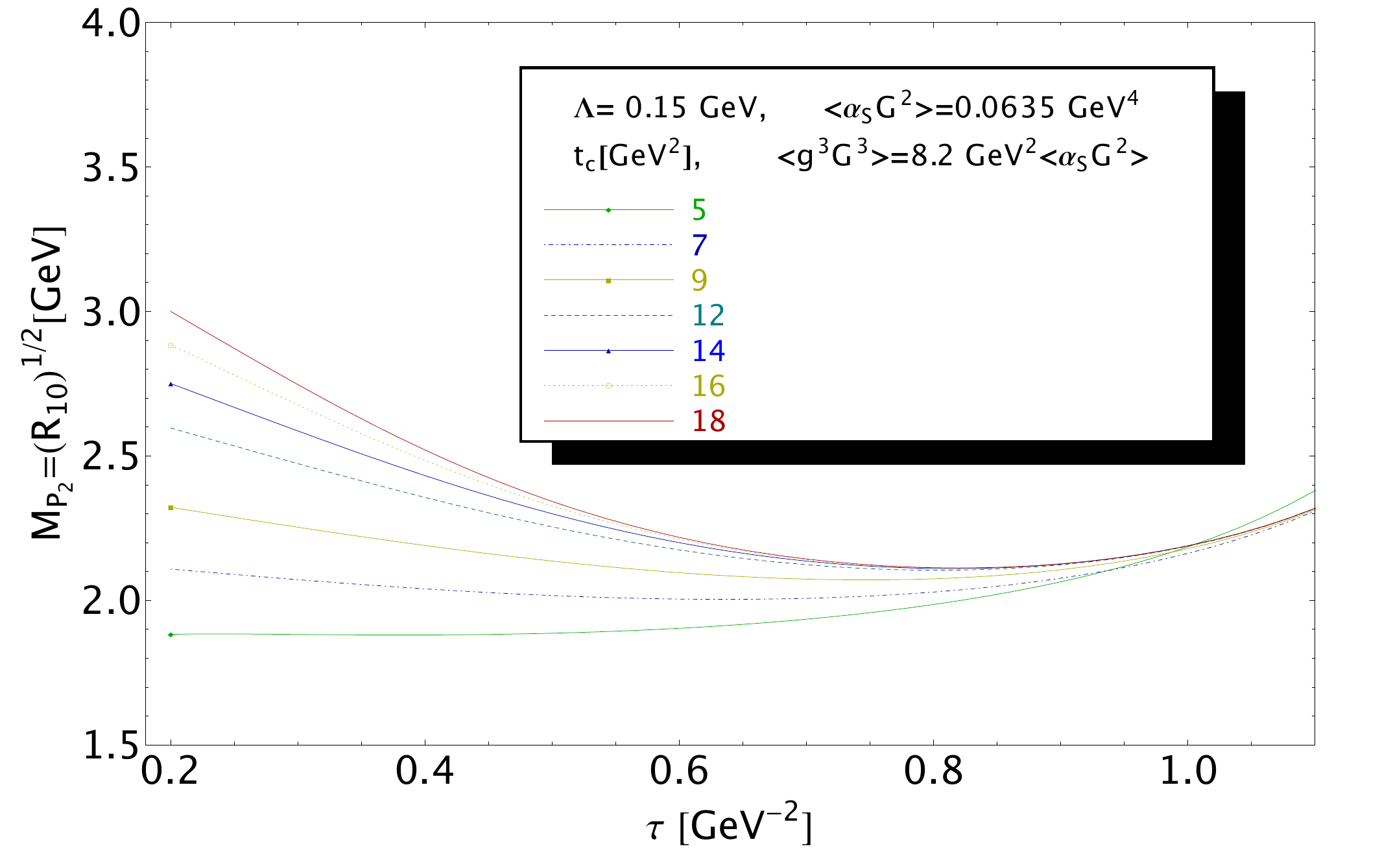} \\
\vspace*{-0.5cm}
\caption{\footnotesize  $M_{P_2}$ from ${\cal R}^c_{10}$  as a function of $\tau$ at N2LO for different values of $t_c$ and for $\Lambda=0.15$ GeV\,\cite{ASNER}.} 
\label{fig:10-steele}
\end{center}
%\vspace*{-0.75cm}
\end{figure} 
%%%%%%%%%%%%%%%%%%%%%%%%%%%%%%%%%%%%%%%
 %%%%%%%%%%%%%%%%%%%%%%%%%%%%%%%%%%%%%%%
\begin{figure}[hbt]
%\vspace*{-0.25cm}
\begin{center}
\includegraphics[width=8.cm]{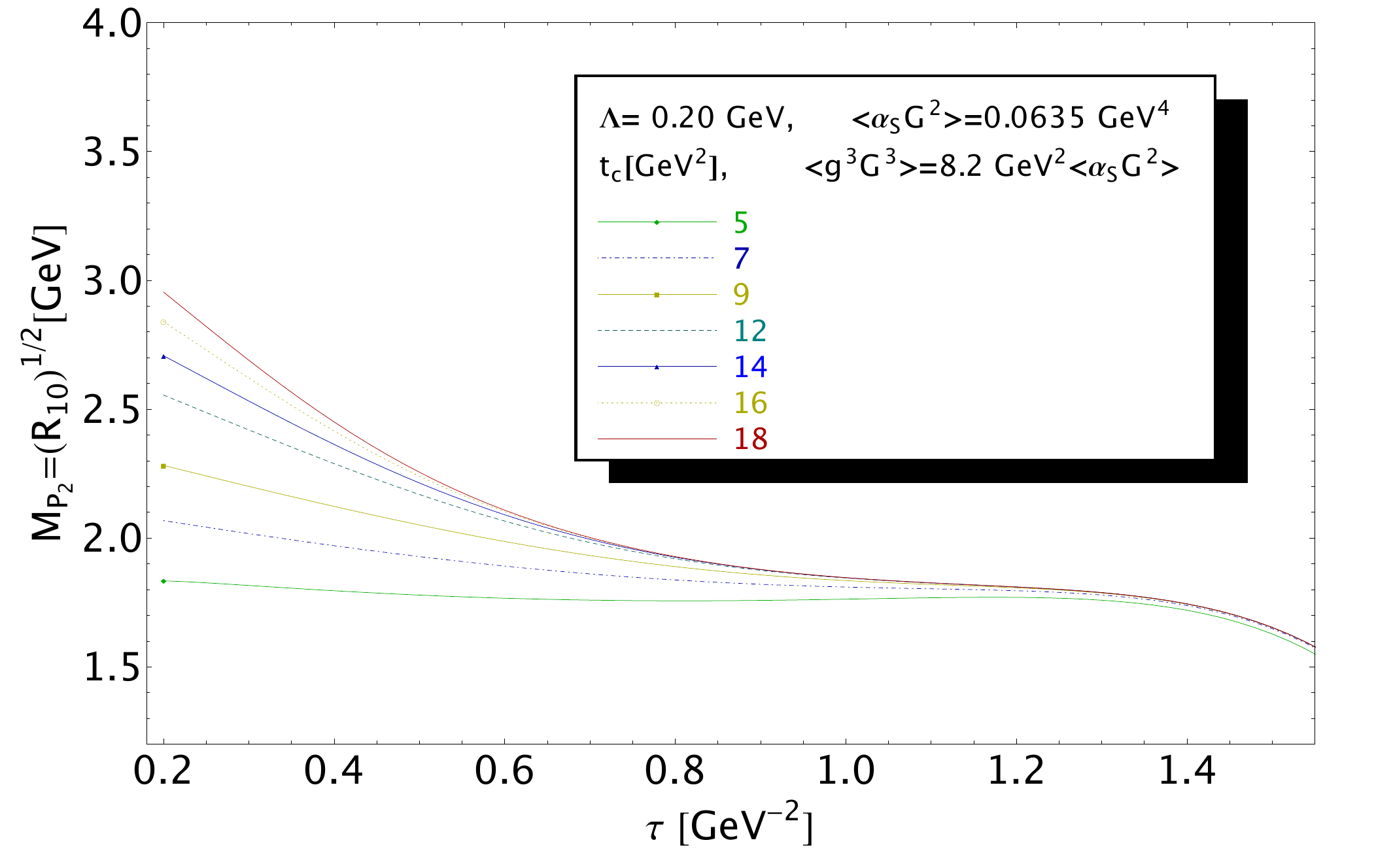} \\
\vspace*{-0.5cm}
\caption{\footnotesize  $M_{P_2}$ from ${\cal R}^c_{10}$  as a function of $\tau$ at N2LO for different values of $t_c$ and for $\Lambda=0.20$ GeV\,\cite{FORKEL}.} 
\label{fig:10-forkel}
\end{center}
%\vspace*{-0.75cm}
\end{figure} 
%%%%%%%%%%%%%%%%%%%%%%%%%%%%%%%%%%%%%%%
 %%%%%%%%%%%%%%%%%%%%%%%%%%%%%%%%%%%%%%%
\begin{figure}[hbt]
%\vspace*{-0.25cm}
\begin{center}
\includegraphics[width=8.cm]{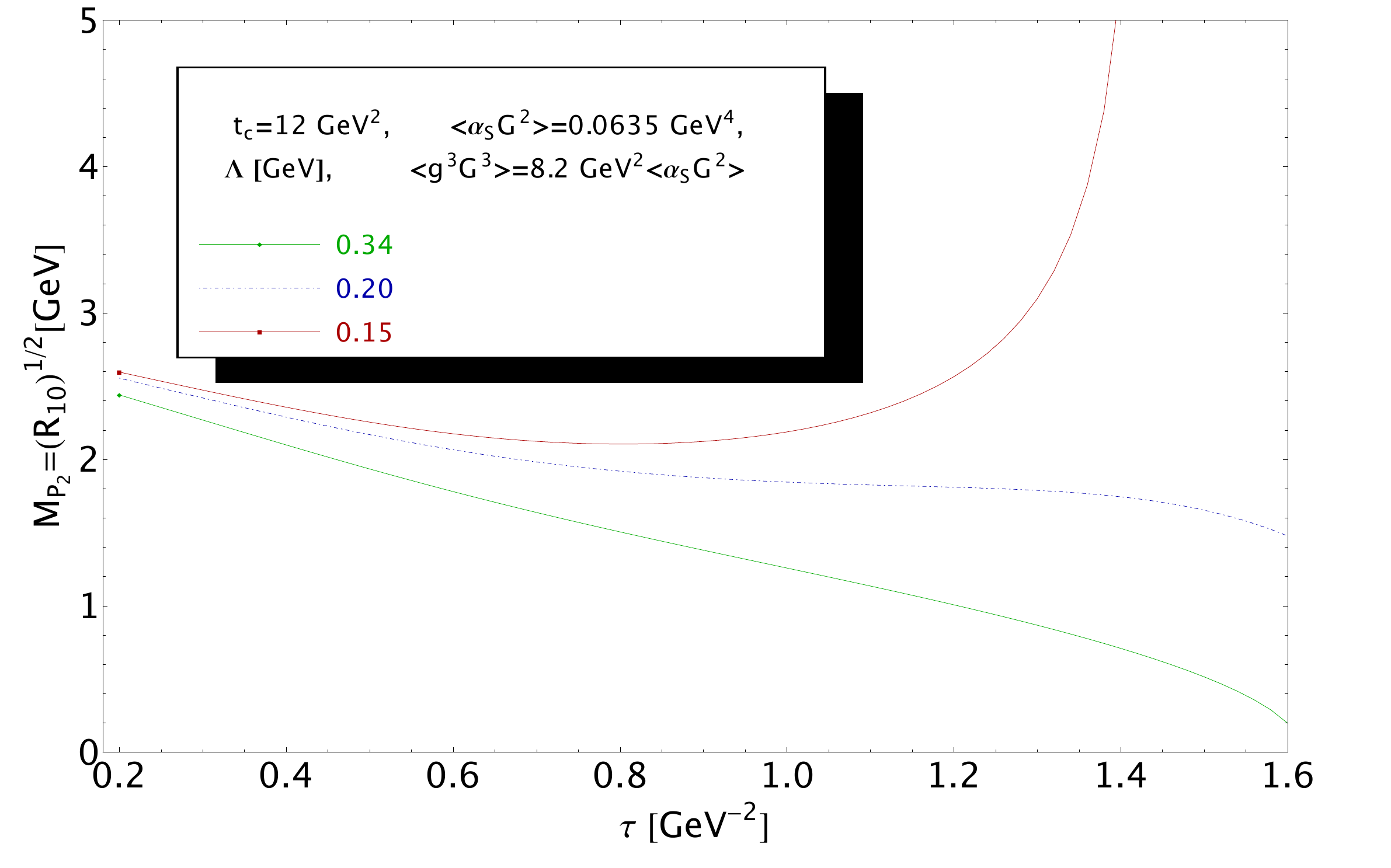} \\
\vspace*{-0.5cm}
\caption{\footnotesize  $M_{P_2}$ from ${\cal R}^c_{10}$  as a function of $\tau$ at N2LO for different values of $\Lambda$ and for $t_c=12$ GeV$^2$.} 
\label{fig:10-Lambda}
\end{center}
%\vspace*{-0.75cm}
\end{figure} 
%%%%%%%%%%%%%%%%%%%%%%%%%%%%%%%%%%%%%%%
\b We update the N2LO results in the literature  using the present values of the QCD parameters given in Table\,\ref{tab:param} without including the $\eta'$-contribution. We show the results of the analysis versus $\tau$ in Fig.\,\ref{fig:10} for different values of $t_c$.  We obtain for  $\Lambda=(340\pm 28)$ MeV :
\beq
M_{P_2}\vert_{\rm N2LO}=  1470(39)_{t_c}(46)_{\tau}(83)_{\Lambda}(44)_{\lambda^2}(19)_{G^2}(11)_{G^3}(31)_{G^4}=1470(118)~{\rm MeV} ,
\label{eq:34}
\eeq
 where we have taken the values  at the inflexion points: $\tau=(0.80\pm 0.04)$ GeV$^{-2}$ and $t_c$ from 7  to 14 GeV$^2$. We have considered the contribution of the tachyonic gluon mass squared $\lambda^2$ as an estimate of the non-calculated higher order terms 
 as expected from\,\cite{CNZa,CNZb,ZAKa,ZAKb}. 
 
\d For $\Lambda = (0.15\pm 0.05) $ GeV, we have a minimum in $\tau$ (Fig.\,\ref{fig:10-steele}). At the minimum, we obtain using the values of the condensates given in Table\,\ref{tab:param} and for $t_c$ from 7  to 14 GeV$^2$ :
\beq
M_{P_2}\vert^{0.15}_{\rm N2LO}=2057(53)_{t_c}(127)_{\Lambda}(44)_{\lambda^2}(14)_{G^2}(43)_{G^3}(70)_{G^4}=2057(166)~{\rm MeV},
\eeq
which reproduces within the error the result of\,\cite{ASNER} quoted in Eq.\,\ref{eq:15} and indicates that the result is quite sensitive to the value of $\Lambda$. The estimate of the HO non-calculated term is quantified by the contribution of $\lambda^2$ in the quoted error. We have not repeated the analysis for the Gaussian sum rules which is expected to reproduce the ressult from LSR at the stability point\,\cite{ZHANG}. 

\d We repeat the analysis for $\Lambda = 0.2 $ GeV, where we have an inflexion point in $\tau$ (Fig.\,\ref{fig:10-forkel}). We  obtain within the standard SVZ expansion and for $\tau=(0.84\pm 0.04)$ GeV$^{-2}$\,:
\beq
M_{P_2}\vert^{0.20}_{\rm N2LO}\simeq 1867(38)_{t_c}(100)_\tau(-)_{\Lambda}(60)_{\lambda^2}(6)_{G^2}(51)_{G^3}(94)_{G^4}=1867(163)~{\rm MeV},
\eeq
indicating that the eventual instanton contribution, with the parameters used in\,\cite{FORKEL},  increases the prediction by about 330 MeV when compared with the one in Eq.\, \ref{eq:20}.

\d We notice that, for a given truncation of the PT series (N2LO here), the shape of the curves of the LSR changes with the value of $\Lambda$ (minimum for $\Lambda$= 0.15 GeV and inflexion points for $\Lambda$= 0,20,\,0.34 GeV). In the same time the value of the mass is very affected by the one of $\Lambda$.

\d In Fig.\,\ref{fig:10-Lambda}, we show the results versus $\tau$ for three different values of $\Lambda$= 0.15, 0.20 and 0.34 GeV and for a fixed value of $t_c$= 12 GeV$^2$. The analysis shows the sensitivity of the results versus $\Lambda$. 

\d Comparing the optimal result for each truncation of the PT series, one has (in unis of MeV):
\beq
M_{P_2}= (2653\pm 327)~ ({\rm LO})~~~ \buildrel -30\%\over \lrar ~~~(1862\pm 18)~ ({\rm NLO})~~~\buildrel -21\%\over\lrar~~~ (1470\pm 39) ~ ({\rm N2LO}),
\label{eq:lo-nlo}
\eeq
where the quoted error comes only from the range of $t_c$-values from 7 to 14 GeV$^2$. One can notice a slow convergence of the result. In the rest of the paper, we re-emphasize that we shall estimate the contribution of the remaining uncalculated higher order (HO) terms of the series from the contribution of the tachyonic gluon mass which we shall add to the errors in the result for $M_P$ and $f_P$.
%%%%%%%%%%%%%%%%%%%%%%%%%%%%%%%%%%%%%%%%%%%%%%%
\section{Gluonia masses from ${\cal R}^c_{mn}$}
%%%%%%%%%%%%%%%%%%%%%%%%%%%%%%%%%%%%%%%%%%%%%%%
\subsection*{\b The analysis}
\begin{figure}[hbt]
%\vspace*{-0.25cm}
\begin{center}
\includegraphics[width=8cm]{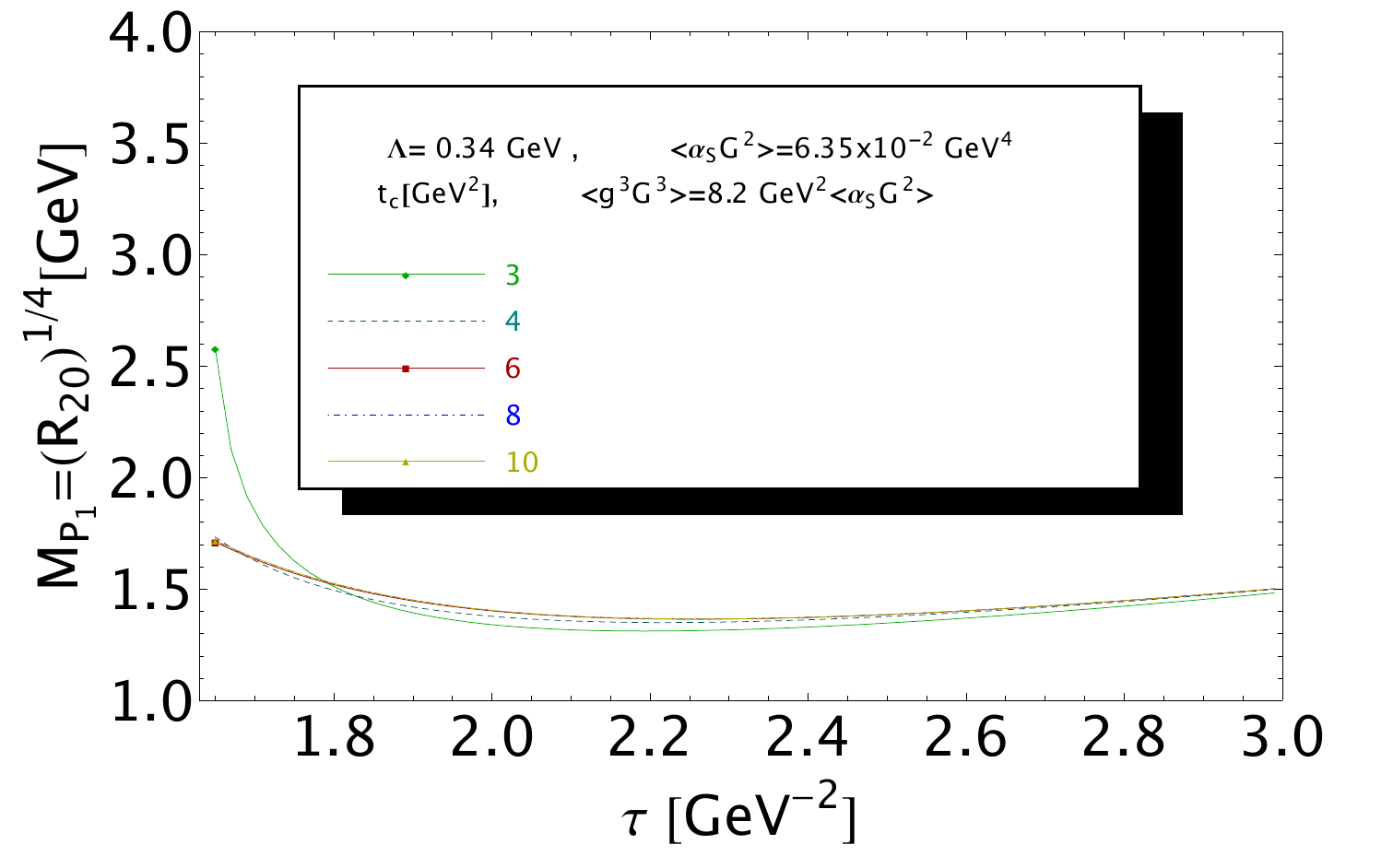} \\
\vspace*{-0.5cm}
\caption{\footnotesize  $M_{P_1}$  from ${\cal R}^c_{20}$  as a function of $\tau$ at N2LO for different values of $t_c$ and for $\Lambda=0.34$ GeV where the $\eta_1$ contribution is included.} 
\label{fig:20}
\end{center}
\vspace*{-0.75cm}
\end{figure}

\begin{figure}[hbt]
%\vspace*{-0.25cm}
\begin{center}
\includegraphics[width=8cm]{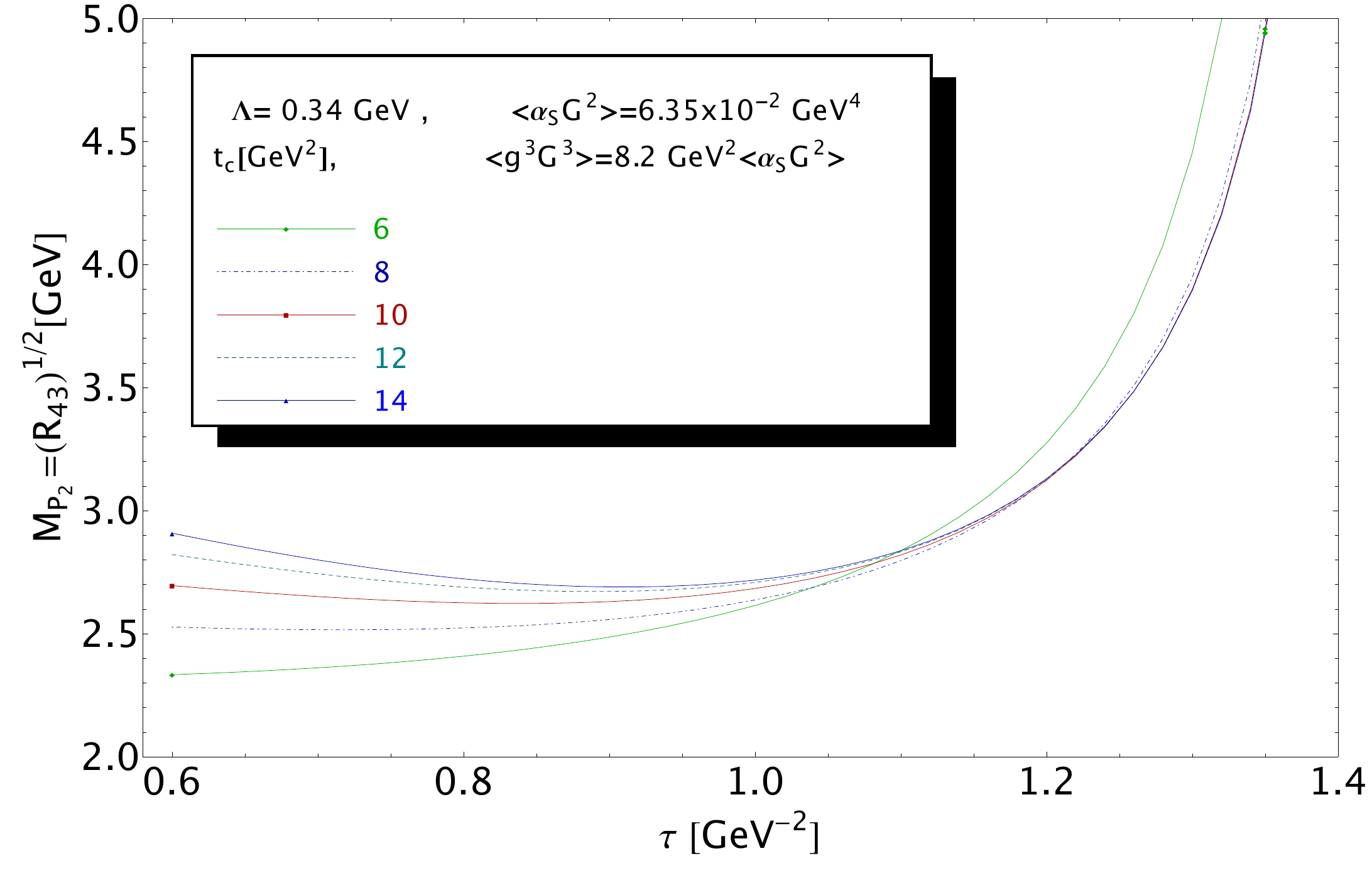} \\
\vspace*{-0.5cm}
\caption{\footnotesize  $M_{P_2}$  from ${\cal R}^c_{43}$  as a function of $\tau$ at N2LO for different values of $t_c$ and for $\Lambda=0.34$ GeV where the $\eta_1\oplus P_1\oplus P'_1$  contributions are included.} 
\label{fig:43}
\end{center}
\vspace*{-0.75cm}
\end{figure}

%%%%%%%%%%%%%%%%%%%%%%%%%%%%
In the following, we shall systematically extract the gluonia masses using different ratios of moments  ${\cal R}^c_{mn}~(n,m=-1,0,...,4)$ within $\tau$- and $t_c$-stability criteria. In order to show explicitly these stability regions specific for a given moment where the optimal results are obtained, we think that is important to show the different figures shown in Fig.\,\ref{fig:20} to \,\ref{fig:43}.  The results are summarized in Table\,\ref{tab:res} and the different sources of errors are given in Table\,\ref{tab:error} where we denote : 

\d  $P_{1,2}$ : One resonance: $P_1$ or $P_2$.

\d $S_{1\eta,2\eta}\equiv P_{1,2}\oplus\eta_1$ : One resonance $\oplus$ $\eta_1$  where $M_{\eta_1}$ and $f_{\eta_1}$ (gluon component of the $\eta'$) are used as inputs from Eqs.\,\ref{eq:meta1} and \,\ref{eq:feta} which have been deduced from previous  sum rules analysis of ${\cal L}_{-1,-2}$.

\d $S_2\equiv P_2\oplus S_{1\eta}$ : Two  resonances ($P_1\oplus P_2) \oplus \eta_1$.

\d $S'_2\equiv P'_1\oplus S_2$ : Three resonances ($P'_1\oplus P_1\oplus P_2)\oplus \eta_1$.

 %%%%%%%%%%%%%%%%%%%%%%%%%%%%%%%%%%%%%%%
%%%%%%%%%%%%%%%%%%%%%%%%%%%%%%%%%%%%%%%

%%%%%%%%%%%%%%%%%%%%%%%%%%%%%%%%%%%%%%%
\begin{figure}[hbt]
%\vspace*{-0.25cm}
\begin{center}
\includegraphics[width=8cm]{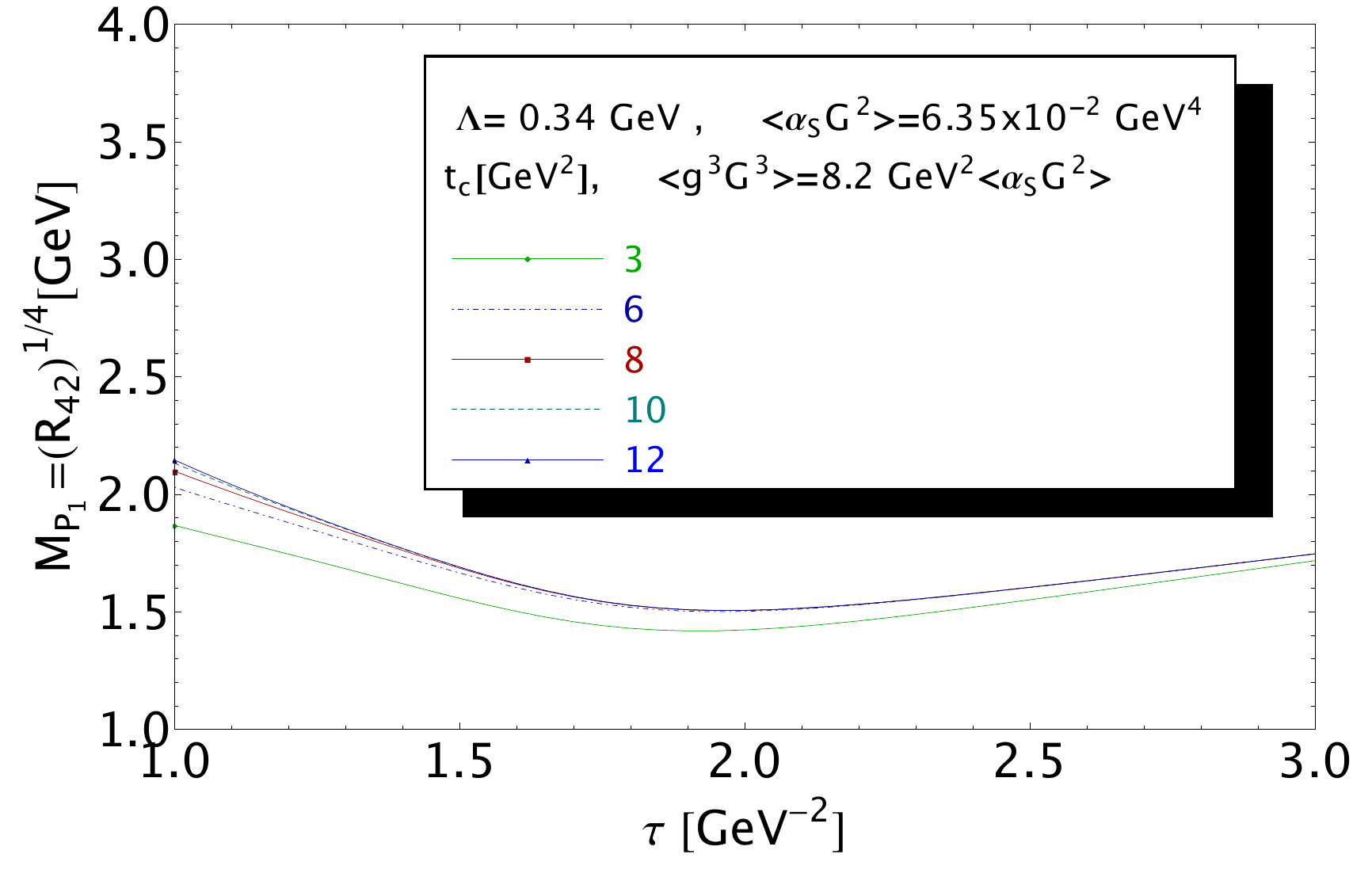} \\
\vspace*{-0.5cm}
\caption{\footnotesize  $M_{P_1}$  from ${\cal R}^c_{42}$  as a function of $\tau$ at N2LO for different values of $t_c$ and for $\Lambda=0.34$ GeV where the $\eta_1$ contribution is included.} 
\label{fig:42}
\end{center}
\vspace*{-0.75cm}
\end{figure} 
%%%%%%%%%%%%%%%%%%%%%%%%%%%%%%%%%%%%%%%
%%%%%%%%%%%%%%%%%%%%%%%%%%%%%%%%%%%%%%%
\begin{figure}[hbt]
%\vspace*{-0.25cm}
\begin{center}
\includegraphics[width=8cm]{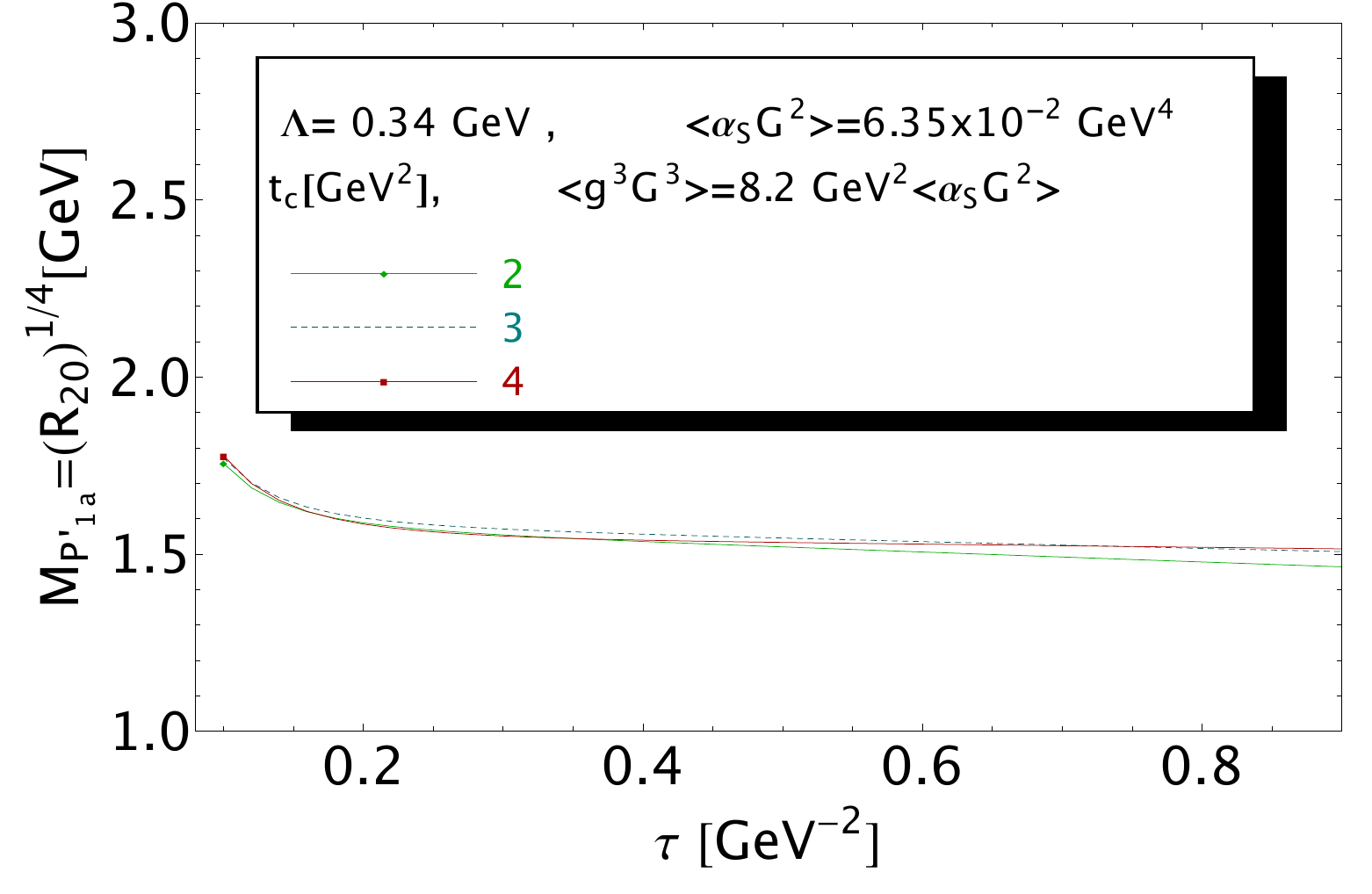} \\
\vspace*{-0.5cm}
\caption{\footnotesize  $M_{P'_{1a}}$  from ${\cal R}^c_{20}$  as a function of $\tau$ at N2LO for different values of $t_c$ and for $\Lambda=0.34$ GeV where the $\eta_1$ and $P_{1a}$ contributions are included.} 
\label{fig:20pprim}
\end{center}
\vspace*{-0.75cm}
\end{figure} 
%%%%%%%%%%%%%%%%%%%%%%%%%%%%%%%%%%%%%%%

%%%%%%%%%%%%%%%%%%%%%%%%%%%%%%%%%%%%%%%
\begin{figure}[hbt]
%\vspace*{-0.25cm}
\begin{center}
\includegraphics[width=8cm]{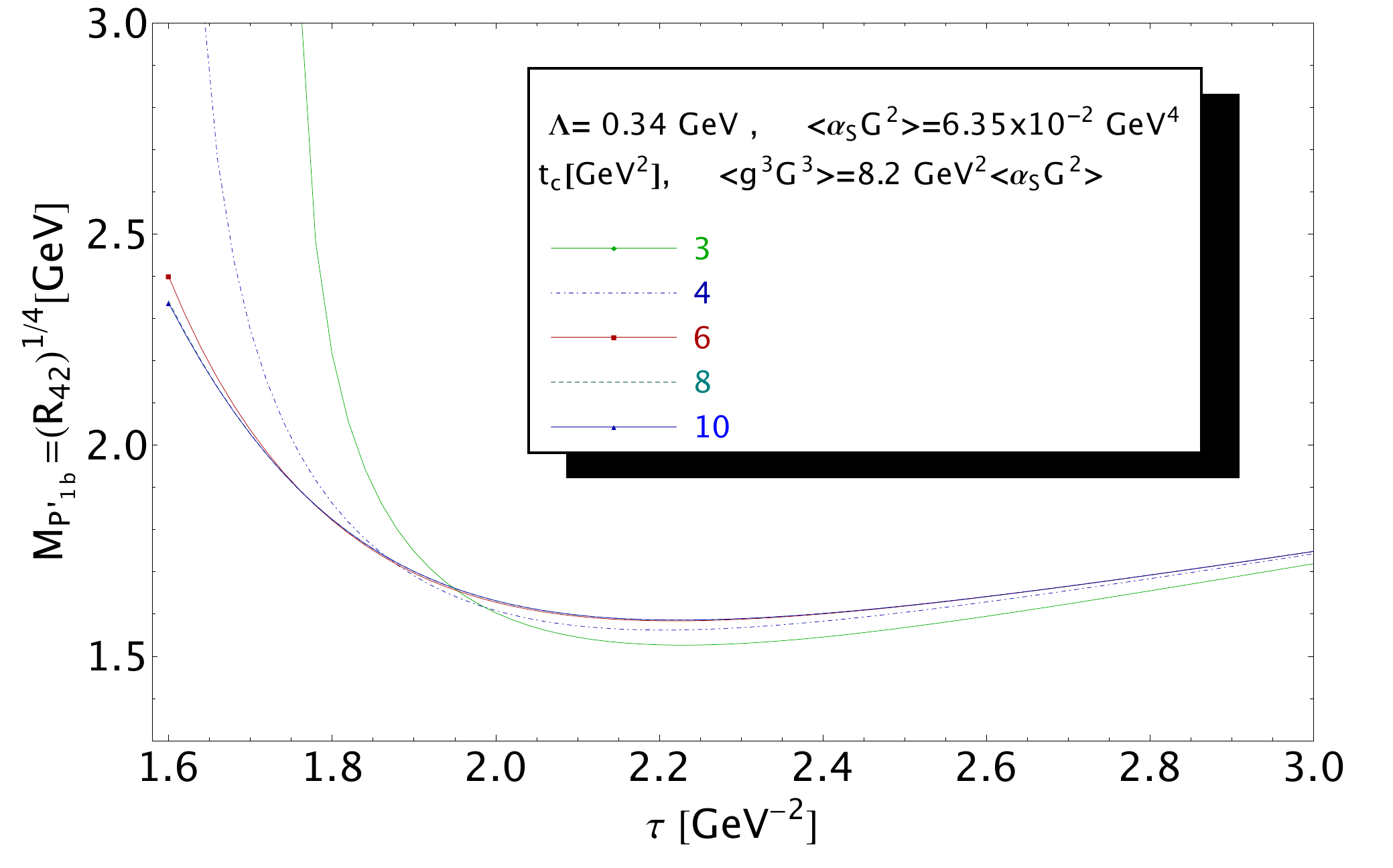} \\
\vspace*{-0.5cm}
\caption{\footnotesize  $M_{P'_{1b}}$  from ${\cal R}^c_{42}$  as a function of $\tau$ at N2LO for different values of $t_c$ and for $\Lambda=0.34$ GeV where the $\eta_1$ and $P_{1b}$ contributions are included.} 
\label{fig:42pprim}
\end{center}
\vspace*{-0.75cm}
\end{figure} 
%%%%%%%%%%%%%%%%%%%%%%%%%%%%%%%%%%%%%%%
%%%%%%%%%%%%%%%%%%%%%%%%%%%%%%%%%%%%%%%%%%%%%%%%%%%%%%%%%%%%%%%%%%%%%%%%%%%%%%%%%%%%
%%%%%%%%%%%%%%%%%%%%%%%%%%%%%%%%%%%%%%%
\begin{figure}[hbt]
%\vspace*{-0.25cm}
\begin{center}
\includegraphics[width=8cm]{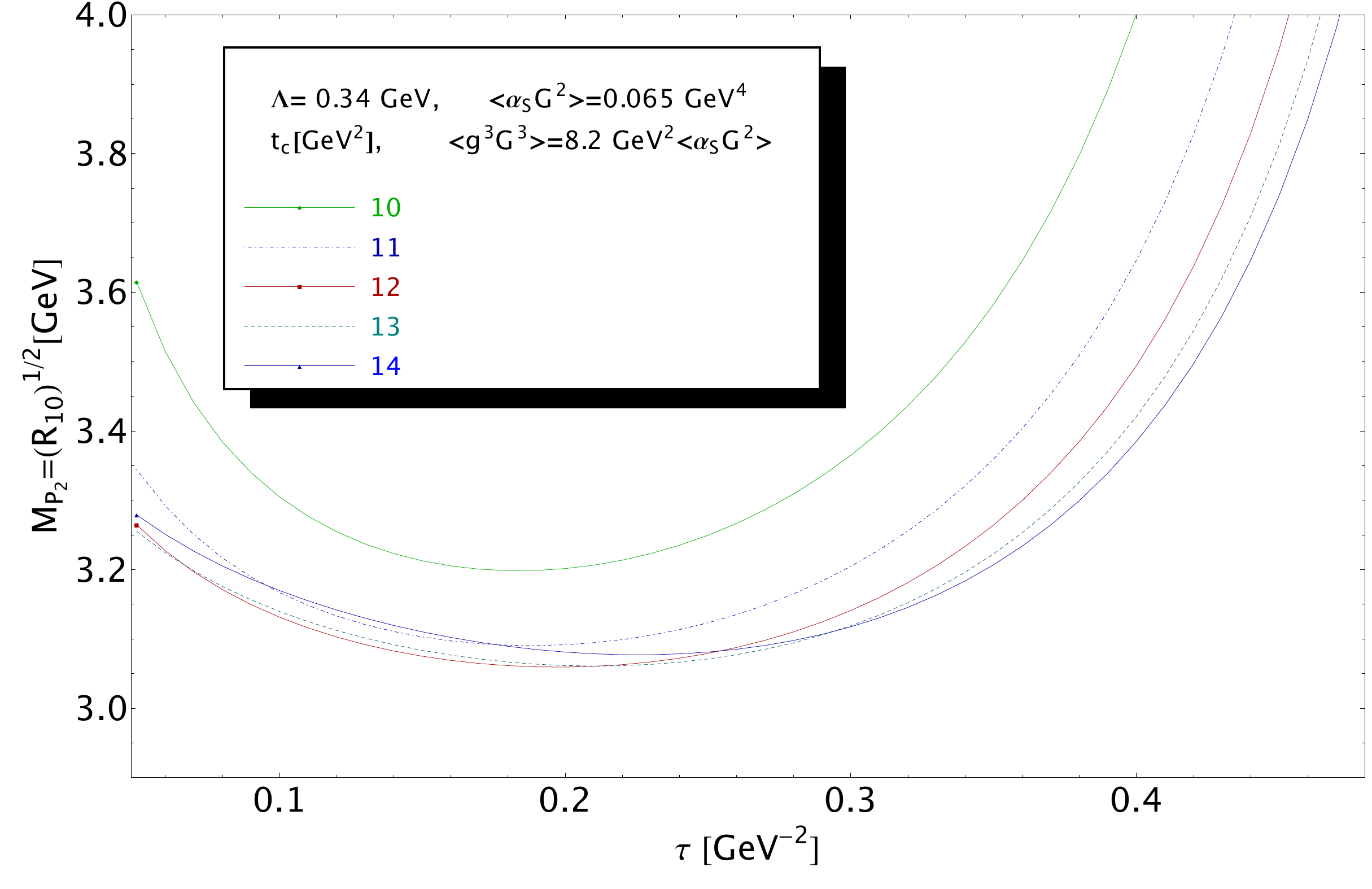} \\
\vspace*{-0.5cm}
\caption{\footnotesize  $M_{P_2}$  from ${\cal R}^c_{10}$  as a function of $\tau$ at N2LO for different values of $t_c$ and for $\Lambda=0.34$ GeV where the $\eta_1\oplus P_1\oplus P'_1$  contributions are included.} 
\label{fig:10b}
\end{center}
\vspace*{-0.75cm}
\end{figure} 
%%%%%%%%%%%%%%%%%%%%%%%%%%%%%%%%%%%%%%%
%%%%%%%%%%%%%%%%%%%%%%%%%%%%%%%%%%%%%%%
\begin{figure}[hbt]
%\vspace*{-0.25cm}
\begin{center}
\includegraphics[width=8cm]{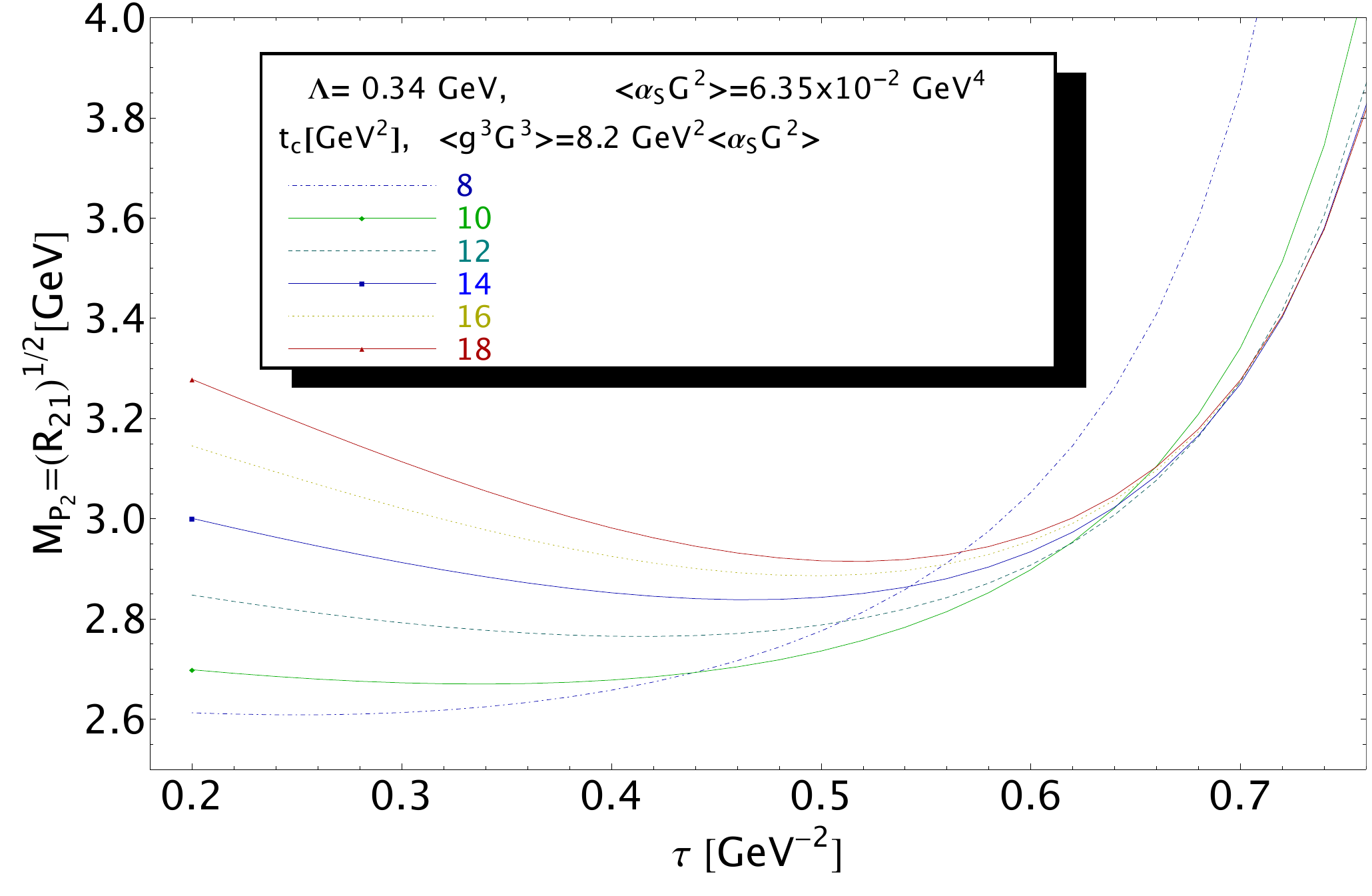} \\
\vspace*{-0.5cm}
\caption{\footnotesize  $M_{P_2}$  from ${\cal R}^c_{21}$  as a function of $\tau$ at N2LO for different values of $t_c$ and for $\Lambda=0.34$ GeV where the $\eta_1\oplus P_1\oplus P'_1$  contributions are included.} 
\label{fig:21}
\end{center}
\vspace*{-0.75cm}
\end{figure} 
%%%%%%%%%%%%%%%%%%%%%%%%%%%%%%%%%%%%%%%
%%%%%%%%%%%%%%%%%%%%%%%%%%%%%%%%%%%%%%%
\begin{figure}[hbt]
%\vspace*{-0.25cm}
\begin{center}
\includegraphics[width=8cm]{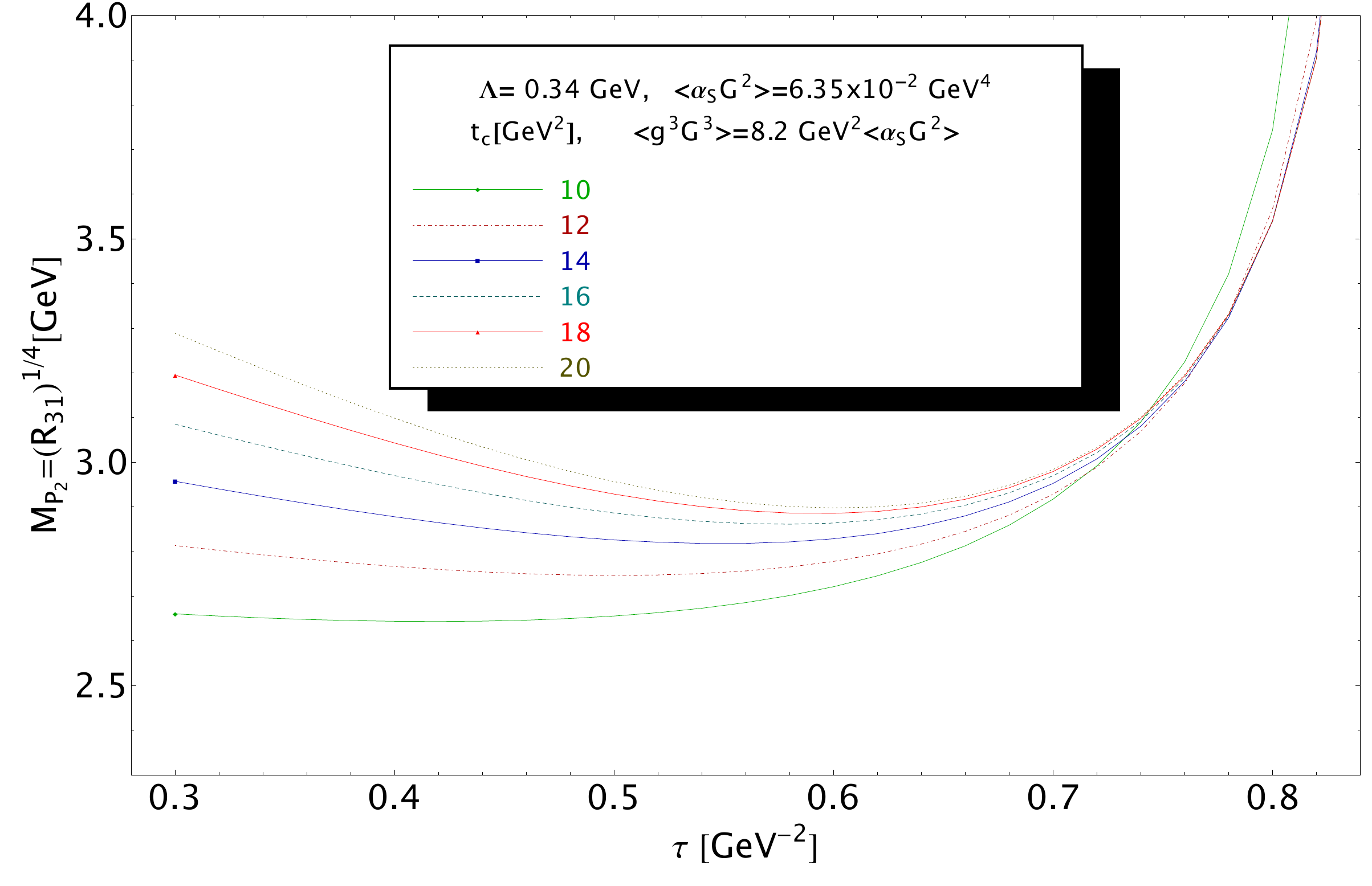} \\
\vspace*{-0.5cm}
\caption{\footnotesize  $M_{P_2}$  from ${\cal R}^c_{31}$  as a function of $\tau$ at N2LO for different values of $t_c$ and for $\Lambda=0.34$ GeV where the $\eta_1\oplus P_1\oplus P'_1$  contributions are included.} 
\label{fig:31}
\end{center}
\vspace*{-0.75cm}
\end{figure} 
%%%%%%%%%%%%%%%%%%%%%%%%%%%%%%%%%%%%%%%
%%%%%%%%%%%%%%%%%%%%%%%%%%%%%%%%%%%%%%%
\begin{figure}[hbt]
%\vspace*{-0.25cm}
\begin{center}
\includegraphics[width=8cm]{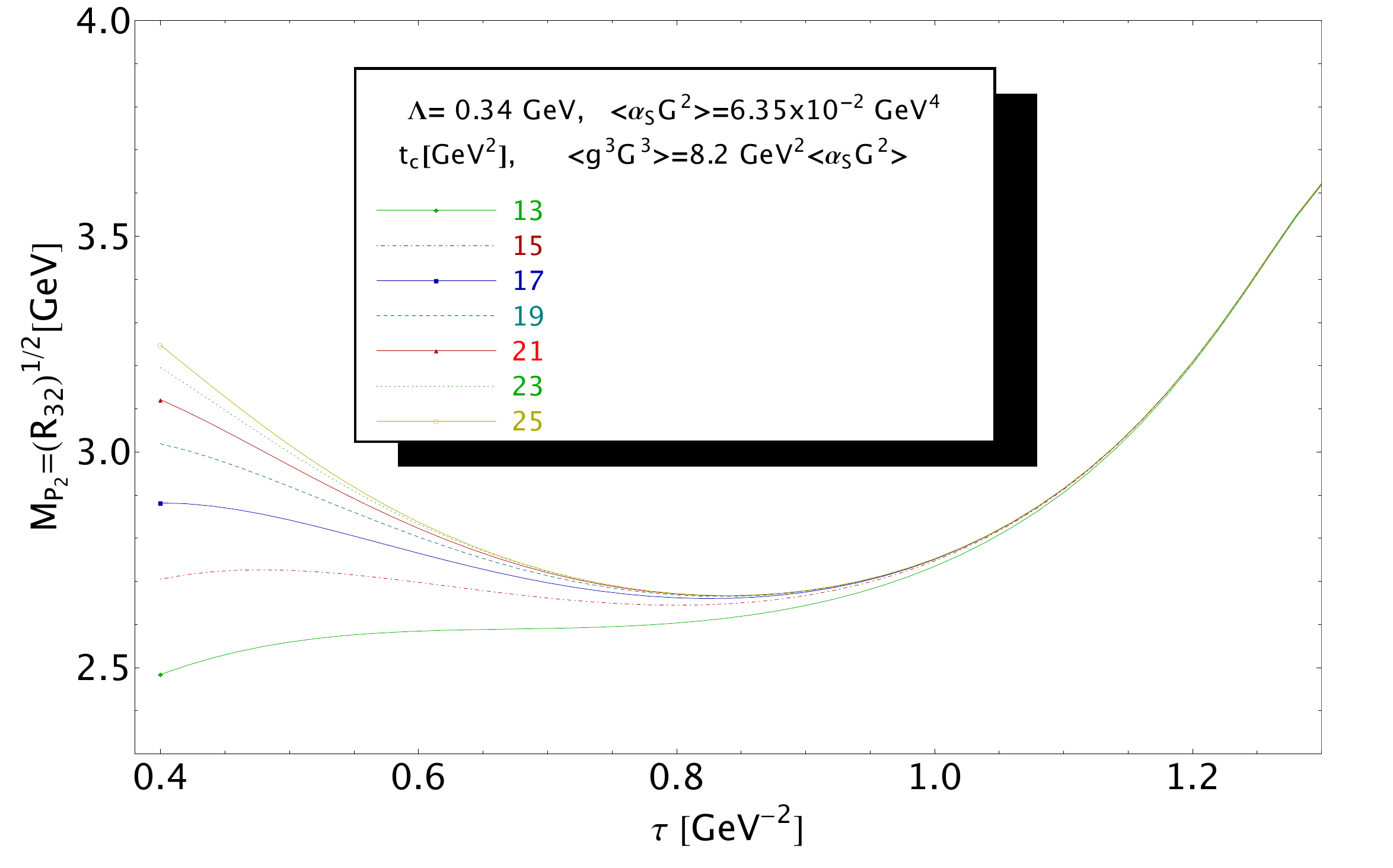} \\
\vspace*{-0.5cm}
\caption{\footnotesize  $M_{P_2}$  from ${\cal R}^c_{32}$  as a function of $\tau$ at N2LO for different values of $t_c$ and for $\Lambda=0.34$ GeV where the $\eta_1\oplus P_1\oplus P'_1$  contributions are included.} 
\label{fig:32}
\end{center}
\vspace*{-0.75cm}
\end{figure} 
%%%%%%%%%%%%%%%%%%%%%%%%%%%%%%%%%%%%%%%

%%%%%%%%%%%%%%%%%%%%%%%%%%%%%%%%%%%%%%%
%%%%%%%%%%%%%%%%%%%%%%%%%%%%%%%%%%%%%%%%

  %%%%%%%%%%%%%%%%%%%%%%%%%%%%%%%%%%%%%
\section{The digluonium decay constants $f_P$}
\begin{figure}[hbt]
%\vspace*{-0.25cm}
\begin{center}
\includegraphics[width=8.5cm]{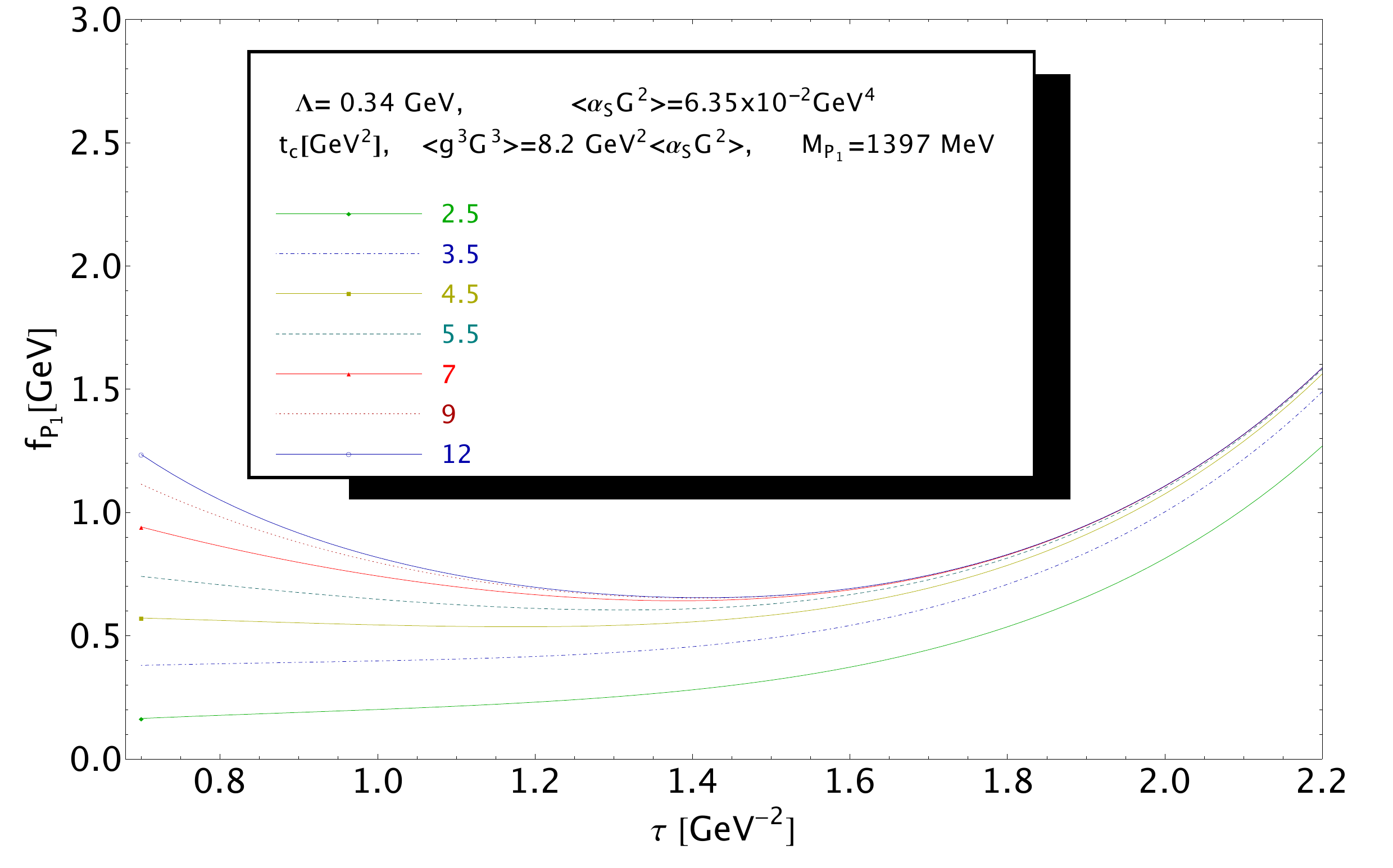} \\
\vspace*{-0.5cm}
\caption{\footnotesize  $f_{P_1}$ from ${\cal L}^c_{2}$  as a function of $\tau$ at N2LO for different values of $t_c$ and for $\Lambda=0.34$ GeV where the $\eta_1$ contribution is included.} 
\label{fig:fp1}
\end{center}
\vspace*{-0.75cm}
\end{figure}

\begin{figure}[hbt]
%\vspace*{-0.25cm}
\begin{center}
\includegraphics[width=8.5cm]{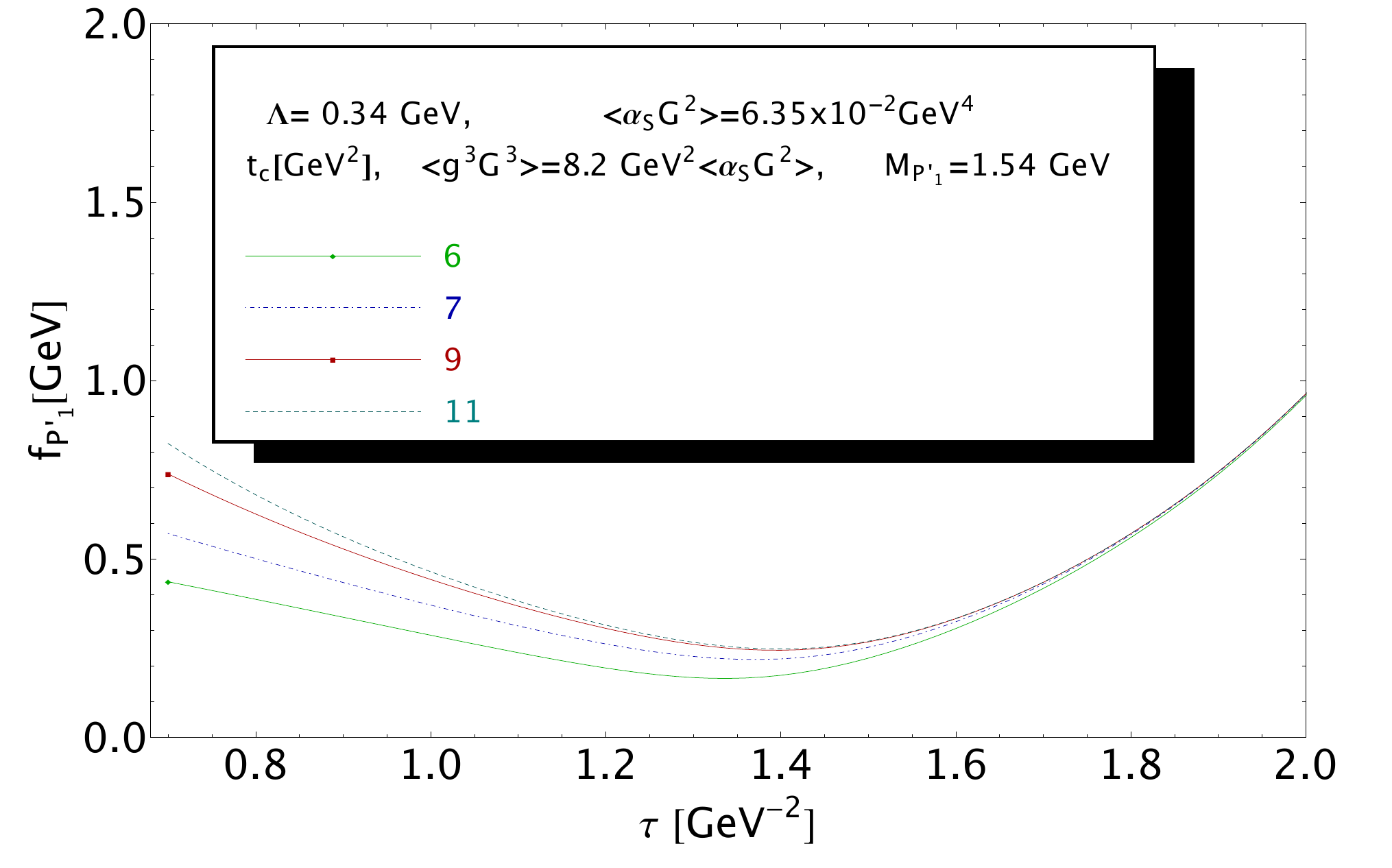} \\
\vspace*{-0.5cm}
\caption{\footnotesize  $f_{P'_1}$ from ${\cal L}^c_{2}$  as a function of $\tau$ at N2LO for different values of $t_c$ and for $\Lambda=0.34$ GeV where the $\eta_1\oplus P_1$ contributions are included.} 
\label{fig:fprim1-2}
\end{center}
\vspace*{-0.75cm}
\end{figure}

\begin{figure}[hbt]
\vspace*{-0.25cm}
\begin{center}
\includegraphics[width=8.cm]{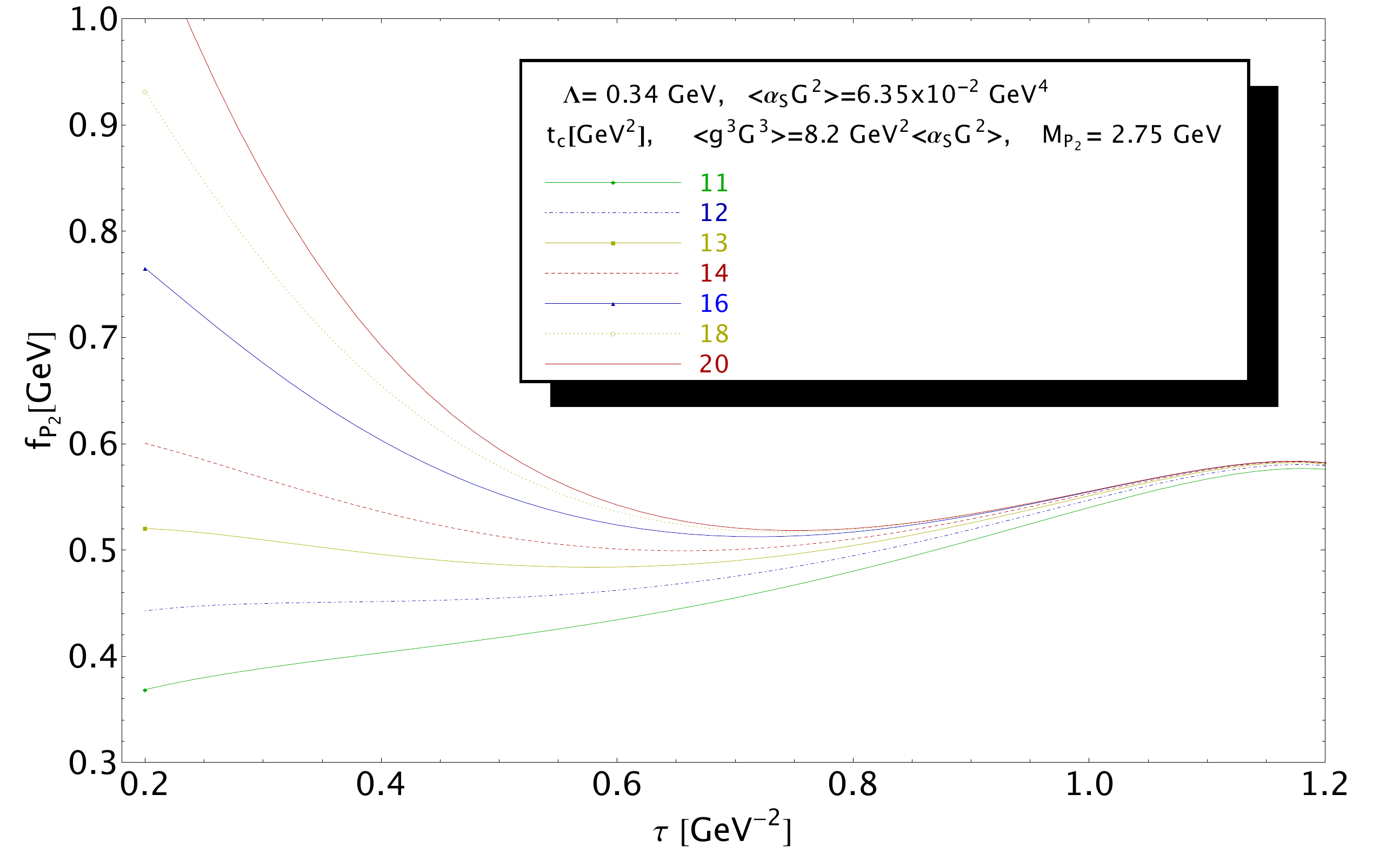} \\
\vspace*{-0.5cm}
\caption{\footnotesize  $f_{P_2}$ from ${\cal L}^c_{3}$  as a function of $\tau$ at N2LO for different values of $t_c$ and for $\Lambda=0.34$ GeV where the $\eta_1\oplus P_1\oplus P'_1$ contributions are included.} 
\label{fig:fp2}
\end{center}
\vspace*{-0.75cm}
\end{figure}

 %%%%%%%%%%%%%%%%%%%%%%%%%%%%%%%%%%%%%
%%%%%%%%%%%%%%%%%%%%%%%%%%%%%%%%%%%%%%%%
The di-gluonium decay constant is defined in Eq.\,\ref{eq:spectral} as $f_\pi=93$ MeV. We shall use the moments ${\cal L}_{2},~{\cal L}_{3}$ (see Figs.\,\ref{fig:fp1} to  \ref{fig:fp2})  for extracting the decay constants of $P_1,~P_2$ and $P'_1$. The results are summarized in Table\,\ref{tab:res} while the different sources of the errors are given inTable\,\ref{tab:error}. 

%%%%%%%%%%%%%%%%%%%%%%%%%%%%%%%%%%%%%%%

%%%%%%%%%%%%%%%%%%%%%%%%%%%%%%%%%%%%%%%
%%%%%%%%%%%%%%%%%%%%%%%%%%%%%%%%%%%%%%%%
%%%%%%%%%%%%%%%%%%%%%%%%%%%%%%%%%%%%%%%
%%%%%%%%%%%%%%%%%%%%%%%%%%%%%%%%%%%%%%%%

%%%%%%%%%%%%%%%%%%%%%%%%%%%%%%%%%%%%%%%%%%%%
\section{Summary of the results}
%%%%%%%%%%%%%%%%%%%%%%%%%%%%%%%%%%%%%%%%%%%%

The results can be grouped into three types of gluonia :\\

\hspace*{0.5cm}\b {\it  Light : $\eta_1$\,\cite{WITTEN,VENEZIANO,DIVECCHIA,SNU1,SHORE}} (see Eqs.\,\ref{eq:meta1} and \,\ref{eq:feta} ):
\beq
M_{\eta_1}=825(45)~{\rm MeV},~~~~~~~~~f_{\eta_1}=905(72)~{\rm MeV}.
\eeq

\hspace*{0.5cm}\b {\it Medium : $P_1$}  and the {\it 1st radial excitations $P'_1$}  (see Table\,\ref{tab:res}):
%\bea
%M_{P_1}&=&1423(87)~{\rm MeV}, ~~~~~~~~~f_{P_1}=491(82)~{\rm MeV},\nnb\\
%M_{P'_1}&=&1581(151)~{\rm MeV}.~~~~~~~~f_{P'_1}=235(133)~{\rm MeV}. 
%\eea
\beq
 [M_{P_{1a}},M_{P_{1b}}]=[1338(112), 1462(117)]~{\rm MeV}, ~~~~~~~~~ [M_{P'_{1a}},M_{P'_{1b}}]=[1508(226), 1553(139)]~{\rm MeV},\label{eq:medium1}
 \eeq
 
 with their mean :
 \beq
 [M_{P_1},f_{P_1}]=[1397(81),594(144)]~{\rm MeV}~,~~~~~~~~~~~~~[M_{P'_1},f_{P'_1}]=[1541(118),205(282)]~{\rm MeV}~,
 \label{eq:medium2}
 \eeq
 
 where the  radial excitation is very closed to the ground state like in the scalar channel\,\cite{SNS21}.  The different sources of the errors are given in Table\,\ref{tab:error}. 
  
  \d Using the positivity of the spectral function, one can also derive a (rigorous) upper bound at the minimum of the ratio of moments ${\cal R}_{20}$ and ${\cal R}_{42}$ (equivalent to take $t_c\to \infty$). Including the $\eta_1$ contribution, it  gives:
 \beq
M_{P_{1a}} \leq  1365(106)~{\rm MeV},~~~~~~~M_{P_{1b}} \leq 1506(104)~{\rm MeV}~~~ \Lrar ~~~M_{P_{1}} \leq 1437(74)~{\rm MeV},
\label{eq:p1bound}
 \eeq 
  where one can notice that in each case, the bound is almost saturated. 
   
 \d One can notice that  the two ratios of moments ${\cal R}^c_{20}$ and ${\cal R}^c_{42}$ stabilize at relatively large value of $\tau\simeq 2$ GeV$^{-2}$\,\footnote{One should mention that ${\cal R}^c_{20}$ present a second minimum at smaller value of $\tau\simeq 0.3$ GeV$^{-2}$ which we shall not consider in this case. For an attempt to get $M_{P_2}$ within a two resonances parametrization, this second minimum disappears.} which make them sensitive to the medium mass $M_{P_1}$ (see Figs.\,\ref{fig:20} and \,\ref{fig:42}) which are not the cases of the other ratios of moments sensitive to the heavy mass $M_{P_2}$ (see the corresponding values of $\tau$ in Table\,\ref{tab:error}).  
 
 \d We check the convergence of the PT series and the OPE at this scale. For a given PT series at N2LO and for different  truncation of the OPE and including the $\eta_1$ contribution, we have for $M_{P_1}$  (in units of MeV):
 \bea
 1154~ {\rm (N2LO)}\lrar1200 ~ ({\rm N2LO}\oplus G^2)\lrar 1683~ ({\rm N2LO} \oplus G^2 \oplus G^3)\lrar 1338({\rm N2LO} \oplus \cdots\oplus G^4)~:~{\cal L}_{20}\nnb\\
 1511~ {\rm (N2LO)}\lrar 1530~ {\rm (N2LO\oplus G^2)} \lrar   1613~ ({\rm N2LO}\oplus G^2\oplus G^3)\lrar 1462({\rm N2LO}\oplus \cdots\oplus G^4)~:~{\cal L}_{42},
 \eea
 and for each truncation of the PT series including the condensates upt to $G^4$:
 \bea
 && 1275~({\rm LO})\lrar 1243 ~ ({\rm NLO})\lrar 1338 ~( {\rm N2LO})~:~{\cal L}_{20}\nnb\\
 && 1298~( {\rm LO})\lrar 1412 ~( {\rm NLO})\lrar 1462 ~({\rm N2LO})~:~{\cal L}_{42},
 \eea
 which show a quite good convergence of the OPE and of the PT series. 
 
 %%%%%%%%%%%%%%%%%%%%%%%%%%%%%%%%%%%%%%%%%%%%%%%
 \hspace*{0.5cm}\b {\it Heavy : $P_2$} (see Table\,\ref{tab:res}):

One can notice from the last column of Table\,\ref{tab:res} that there is a splitting of about 400 MeV for $M_{P_2}$=3059(385) MeV from ${\cal R}_{10}^c$ and the other $P_2$ states which  may indicate that there can be more than one state in the region above 2 GeV.  However, as the error is large  we cannot strictly confirm this observation and consider the  conservative mean value of $M_{P_2}$ from all determinations:
 \beq
M_{P_2}= 2751(140)~{\rm MeV}, ~~~~~~~~~~~~~f_{P_2}=500(43)~{\rm MeV},
\label{eq:mp2}
\eeq
where the errors are mainly due to $\Lambda,~f_{P_1}$ and $M_{P_1}$ (see Table\,\ref{tab:error}). 
 %%%%%%%%%%%%%%%%%%%%%%%%%%%%%%%
  %%%%%%%%%%%%%%%%%%%%%%%%%% 
  \vspace*{-0.25cm}
\section{Comparison with some existing estimates and confrontation with the data}
%%%%%%%%%%%%%%%%%%%%%%%%%%%
\subsection*{\b  Previous QCD spectral sum rules}
%%%%%%%%%%%%%%%%%%%%%%%%%%%
%%%%%%%%%%%%%%%%%%%%%%%%%%%%%%%%
-- We have already commented some results obtained at N2LO in Section\,\ref{sec:r10} where we have emphasized the large effect of $\Lambda$ QCD on the results. Using the most recent value of $\Lambda$, the N2LO results from\,\cite{ASNER,FORKEL,ZHANG} working with the ${\cal R}_{10}$ ratio of moments  contribution become $M_{P_2}=1470(101)$ MeV for ``one resonance" and $M_{P_2}=2097(264)$ MeV for  ``one resonance $\oplus\,\eta_1$" (see Table\,\ref{tab:res}). 

-- Comparing the LO result of\,\cite{ZHU} from ${\cal R}_{10}$ within a ``One resonance $\oplus $ QCD continuum'' parametrization of the spectral function with the one in Eq.\,\ref{eq:lo-nlo}, one can see that the two LO results agree wthin the errors though the one of \,\cite{ZHU} is obtained outside the $\tau$-stability region. The apparent agreeement of this LO result for ``one resonance" from ${\cal R}_{10}$ with the N2LO one with multiple resonances from the same ${\cal R}_{10}$ is quite lucky due to the large effect of radiative corrections which decreases the LO result by a large amount for this moment as discussed previously.  

%%%%%%%%%%%%%%%%%%%%%%%%%%%
\subsection*{\b  Lattice calculations}
%%%%%%%%%%%%%%%%%%%%%%%%%%%

\d Lattice results are in the range (2.15 $\sim$ 2.72) GeV\,\cite{RAGO,MEYER,MATHIEU,CHEN} which can be compared with the value of $M_{P_2}=2639(127)$ MeV obtained in Eq.\,\ref{eq:mp2} and Table\,\ref{tab:res}.  However, a direct comparison of the results from the two approaches cannot be done properly as the lattice does not detect the medium gluonia in Eqs.\,\ref{eq:medium1} and \ref{eq:medium2} though the decay constant $f_{P_1}$ is almost equal to $f_{P_2}$ (see Table\,\ref{tab:res}).  

\d From the present approach, this feature is
due to the fact that the resonance contributes to   ${\cal L}_n$ as:
\beq
 {\cal L}_n\simeq \sum_{i=1,\cdots} 2f^2_{P_i}M^{2(n+2)}_{P_i}\,e^{-M_{P_i}^2\tau},
 \eeq
 where the exponential factor is expected to kill the high-mass resonance contributions to the sum rule moments (analogous exponetial factor plays a similar role  in the lattice calculations). 
Taking $M_{P_1,P_2}$ and $f_{P_1}\approx f_{P_1,P_2}$ in Table\,\ref{tab:res} and $\tau\approx 0.5$ GeV$^{-2}$ for the moments used to get $M_{P_2}$, it is easy to check that the suppression due to the exponential weight is not enough to enhance the lowest ground state contribution, where the ratio behaves as:
\beq
\frac{{\cal L}_n(P_2)}{{\cal L}_n(P_1)}\approx 0.96 \ga \frac{M_{P_2}}{M_{P_1}}\dr^{2n}~:~~~~~ n=1,2,...,
\eeq
indicating that working with the high moments which stabilize at low values of $\tau$, one can miss the lowest mass $P_1$. The opposite situation is obtained for the moments used to get $M_{P_1}$ which stabilizes at larger $\tau\simeq 2$ GeV$^{-2}$ enabling to extract $M_{P_1}$ where the $P_2$ contribution is negligible.
%%%%%%%%%%%%%%%%%%%%%%%%%%%
\subsection*{\b  Some other approaches}
%%%%%%%%%%%%%%%%%%%%%%%%%%%
\d Holographic approach gives a mass of about 2.1 Gev\,\cite{HOLO}.

\d In contrast to the previous approaches, a recent inverse problem dispersive method found a mass of 1750 MeV\,\cite{HSIANG}, while the flux tube and mixing matrix models predict  1.4 GeV\,\cite{FADEEV,LIU} which is in the range of   our mean medium mass  predictionsw $M_{P_1}=1397(81)$ MeV and  $M_{P'_1}=1541(118)$ MeV in Eq.\,\ref{eq:medium2}.

%%%%%%%%%%%%%%%%%%%%%%%%%%%
\subsection*{\b Confrontation with the data below 2 GeV}
%%%%%%%%%%%%%%%%%%%%%%%%%%%
\d We expect that the new medium gluonia $P_1(1397)$ [or $P_{1a}(1338)$ and $P_{1b}(1462)$] and their first radial excitations [$P'_1(1541)$ (or $P'_{1a}(1508)$ and $P'_{1b}(1553)$] obtained in this paper will bring some light for a much better understanding of the  $0^{-+}$ $\eta$-like states found below 2 GeV.
  
     \d $P_1(1397)$ supports the experimental facts that the $\eta(1405)$ is an excellent gluonium candidate. If we assume that the $P_{1a}$ and $P_{1b}$ are different states, we may expect than $P_{1a}$ brings a small gluon component to the $\eta(1295)$ explaining its seen decay to $\eta(\pi\pi)_{S-wave}$ via (most probably) the $\sigma$ which is expected to be the lighest scalar gluonium\,\cite{SNS21,VENEZIA,SNG}, while $P_{1b}$ can explain the gluonium nature of the $\eta(1405)$ which is also seen to decay into $\eta(\pi\pi)_{S-wave}$. 

   \d $P'_1(1541)$ may suggest that the $\eta(1495)$ can possess a gluon component though smaller than the one of the ground state $P_1(1397)$ due to its weaker coupling to the gluonic current: $f_{P'_1}\leq f_{P_1}$. If $P'_1$ is splitted into $P'_{1a}$ and $P'_{1b}$, then, $P'_{1a}$ and $P'_{1b}$ may share some gluons to $\eta(1495)$ and $\eta(1760)$. 
%%%%%%%%%%%%%%%%%%%%%%%%%%%%%%%
 \vspace*{-0.25cm}
 \section{Summary and Conclusions}
 %%%%%%%%%%%%%%%%%%%%%%%%%%%%%%%
%%%%%%%%%%%%%%%%%%%%%%%%%%%%%%%%%%%%%%%%%%%
 For an attempt to study the topics adressed in the title of this paper, we have scrutinized and improved the $0^{-+}$ pseudoscalar gluonium sum rules using updated values of the QCD input parameters and using a multiple resonance parametrization of the spectral function beyond the minimal duality ansatz : ``one resonance $\oplus$ QCD continuum" and high degree LSR moments 

 %%%%%%%%%%%%%%%%%%%%%%%%%%%%%%%%%%
 \subsection*{\b Slope of the Topological Charge and Proton Spin}
 %%%%%%%%%%%%%%%%%%%%%%%%%%%%%%%%%%
 \d The properties of the $\eta_1$ singlet piece of the $\eta'$ as well as the value of the topological charge $\chi(0)$ from $U(1)_A$ large $N_c$ and current algebra approaches are reproduced from the approach\,\cite{SNU1,SNU2,SNG0,SNG1,SHORE}.
 
  \d  We have estimated the slope of the topological charge at N2LO and find it to be  $\sqrt{\chi'(0)}(Q^2=2\,\rm GeV^2)=24.3(3.1)$ MeV  (Eq.\,\ref{eq:chiprim}). It confirms the previous NLO value of 22.3(4.8) MeV\,\cite{SHORE} from LSR updated here, which  is  still smaller than the OZI value $f_\pi/\sqrt{6}= 38$ MeV. 

 \d We have used this value of  $\sqrt{\chi'(0)}$ to estimate the singlet form factor $G_A^{(0)}$ of the axial current \,(see Eq.\ref{eq:ga}) and the first moment $\int_0^1\, dx\, g_1^P(x)$ of the polarized proton structure function\,(see Eq.\ref{eq:spin1}) which remarkably agree with the data\,\cite{EMC,COMPASS,HERMES}. 

  %%%%%%%%%%%%%%%%%%%%%%%%%%%%%%%%%%
 \subsection*{\b $0^{-+}$ Pseudoscalar Gluonia}
 %%%%%%%%%%%%%%%%%%%%%%%%%%%%%%%%%%
 We have used more than one resonances and worked with different high degree moments to analyze the complex spectra of the observed $\eta$-like gluonia. We find that\,:  

 \d  The lattice results for the pseudoscalar gluonium mass are comparable with the so-called heavy gluonium mass $P_2(2751)$ obtained here from moments ratios which stabilize at smaller values of $\tau$. 

 \d New gluonia $P_1$-like states with medium masses are found from some ratios of moments which stabilize at larger values of $\tau$ and can eventually  explain the gluonium nature of some of the four $\eta$-like states found below 2 GeV. Our result supports the gluonium nature of the experimental candidate $\eta(1405)$  and suggests some gluon component to the other $\eta$-like wave functions. 
 
 \d It is also remarkable to notice that the structure of the pseudoscalar gluonia spectrum is very similar to the one of the scalar gluonia (its chiral partner) studied recently in\,\cite{SNS21}.  There is a (natural) one to one correspondence between the pseudoscalar gluonia and their chiral scalar analogue from Ref.\,\cite{SNS21}: $\sigma(1)\to \eta_1;~G_1(1.55)\to P_{1}; ~\sigma'(1.1),G'(1.56)\to P'_{1a,1b};~G_2(3)\to P_2$ which is mainly due to the importance  of the QCD PT contributions  that are almost equal in the analysis of these two channels.

 %%%%%%%%%%%%%%%%%%%%%%%%%%%%%%%%%%%%%%%%

\begin{table}[H]
\setlength{\tabcolsep}{0.3pc}
% -----------------------------------------------------
% adapted from TeX book, p. 241
%\newlength{\digitwidth} \settowidth{\digitwidth}{\rm 0}
\catcode`?=\active \def?{\kern\digitwidth}
% -----------------------------------------------------
%\label{tab:res0}
\begin{center}
\vspace*{-0.25cm}
    {\scriptsize
  \begin{tabular}{lllllllll}
\hline

%{ \bf  Observables }&\bf LSR&\normalsize\bf Fig.\#&\bf 1 resonance&\bf 1 resonance \boldmath$\oplus~\eta_1$&\bf 2 resonances \boldmath$\oplus~\eta_1$\\
{Observables }& LSR& Fig.\#&$\eta_1$& $P_{1,2}$& $S_{1\eta,2\eta}\equiv P_{1,2}$ $\oplus~\eta_1$&  $S_2\equiv P_2~\oplus~S_{1\eta}$&$S'_{1} \equiv P'_1~\oplus~S_{1\eta}$& $S'_2\equiv P_2~\oplus~S'_1 $\\

% \\
\hline
%\hline
%\\
%{\boldmath  $P_1$}\\
%{$M_{P_1}$\, [MeV]} 
%\cline{0-0} 
%\\
%{\it Ground states}\\
{\it $M_P$  [MeV] }\\
\cline{0-0}
  $P_1$&${\cal R}^c_{20}$&\ref{fig:20}&&1205(126)&1338(112)&\\
  --  &--&&&&$\leq 1365(106)$&\\
--&${\cal R}^c_{42}$&\ref{fig:42}&&1432(115)&1462(117)& \\
 --  &--&&&&$\leq 1506(104)$&\\
--&\it Mean&&&\it 1345(83)&\it 1397(81)&\\
 --   &--&&&&$\leq\it 1437(74)$&\\
$P'_{1a}$&${\cal R}^c_{20}$&\ref{fig:20pprim}&&&&&1508(226)\\
$P'_{1b}$&${\cal R}^c_{42}$&\ref{fig:42pprim}&&&&&1553(139)\\
$P'_{1}$&\it Mean&&&&&&\it 1541(118)&\\
$P_2$
&${\cal R}^c_{10}$&\ref{fig:10b}&&1470(101)&2097(264)&2990 (318)&&3059(385)\\
--& ${\cal R}^c_{21}$&\ref{fig:21}&&1846(73)&2005(119)&2715(285)&&2778(340)\\
--&${\cal R}^c_{31}$&\ref{fig:31}&&1726(71)&2050(119)&2699(265)&&2764(317)\\
--&${\cal R}^c_{32}$&\ref{fig:32}&&2096(44)&2261(39)&2534(327)&&2644(391) \\
--&${\cal R}^c_{43}$&\ref{fig:43}&&2054(42)&2080(158)&2620(205)&&2661(227) \\
--&\it Mean&&&\it 1965(25)&\it 2214(34)&\it 2694(119)&&\it 2751(140)\\
%--&${\cal R}^c_{43}$&&&&&&&$\leq 2636(165)$\\
%\\
\cline{0-0}
\it $f_{P}$  [MeV]\\
\cline{0-0}
$\eta_1$&${\cal L}^c_{-1}$&\ref{fig:feta1}&$905(72)$\\
%$P_1$&${\cal L}^c_{-1}$&\ref{fig:fpm1}&&&435(103)&\\
$P_1$&${\cal L}^c_{2}$&\ref{fig:fp1}&&&$594(144)$& \\
%--&\it Mean&&&&\it 494(82)& \\
$P'_1$&${\cal L}^c_{2}$&\ref{fig:fprim1-2}&&&&&$205(282)$& \\
%&${\cal L}^c_{3}$&\ref{fig:fprim1-3}&&&&&286(167)& \\
%&\it Mean &&&&&\it 287(132)&\\
$P_2$&${\cal L}^c_{3}$&\ref{fig:fp2}&&&&$504(39)$&&$500(43) $\\
%\\
 %\\
\hline
%\hline
\end{tabular}
{\scriptsize
 \caption{Pseudoscalar di-gluonium masses from LSR for PT at N2LO and gluon condensates at NLO up to $D=8$ for : ``$\sum_{P_i}$ resonances $\oplus$ QCD continuum"  parametrization of the spectral function.  The mean value of $M_{P'_1}$ is used to get $f_{P'_1}$ and the one $[M_{P'_1},f_{P'_1}]$  for $[M_{P_2},f_{P_2}]$ in the last column. 
 The bounds on $M_{P_1}$ in column \# 4 are obtained at the minmum of ${\cal R}_{20,42}$ using the positivity of the spectral function and  including the $\eta_1$ contribution. The different sources of errors are given in Table\,\ref{tab:error}.
 %The one of $M_{P_2}$ in the last column comes from  ${\cal R}_{43}$ after including  $\eta_1, P_1$ and $P'_1$. $M_{\eta_1},f_{\eta_1}$ in Eqs.\,\ref{eq:meta1}  \,\ref{eq:feta1}
 The QCD parameters in Table\,\ref{tab:param} have been used as inputs.}
 \label{tab:res}
 \vspace*{-0.5cm}
}
}
\end{center}
\end{table}
%%%%%%%%%%%%%%%%%%%%%%%%%%%%%%%%%%%%%%%%%%%%
  %%%%%%%%%%%%%%%%%%%%%%%%%%%%%%%%%%%%%%%%%%%
%{\tiny
\begin{table}[H]
 \vspace*{-0.5cm}
\setlength{\tabcolsep}{0.2pc}
 \begin{center}
{\scriptsize{
\begin{tabular}{lllllllllllllll llllll}
%{\textwidth}{ll ll  ll  ll ll ll ll ll ll r{@{}ll ll  ll  ll ll ll ll ll ll r@{\extracolsep{\fill}}l}
%\hline
\hline
                 &\# Resonances& LSR
                				&\multicolumn{1}{c}{$ t_c $[GeV$^2$]}
                				&\multicolumn{1}{c}{$\Delta t_c$}
					&\multicolumn{1}{c}{$\tau$[GeV$^{-2}$]}
					&\multicolumn{1}{c}{$\Delta \tau$}
					%&\multicolumn{1}{c}{$\Delta \mu$}
					&\multicolumn{1}{c}{$\Delta \Lambda$}
					%&\multicolumn{1}{c}{$\Delta HO$ } 	
					&\multicolumn{1}{c}{$\Delta \lambda^2$ } 						                          &\multicolumn{1}{c}{$\Delta G^2$}
								
					&\multicolumn{1}{c}{$\Delta G^3$}
					&\multicolumn{1}{c}{$\Delta G^4$}
					&\multicolumn{1}{c}{$\Delta f_{\eta_1}$}
					&\multicolumn{1}{c}{$\Delta M_{\eta_1}$}
					%&\multicolumn{1}{c}{$\Delta M_{(G)_1}$}
					&\multicolumn{1}{l}{$\Delta M_{P_1}$}
					&\multicolumn{1}{l}{$\Delta f_{P_1}$}
					&\multicolumn{1}{l}{$\Delta M_{P'_1}$}
					&\multicolumn{1}{l}{$\Delta f_{P'_1}$}
					&\multicolumn{1}{l}{$\Delta M_{P_2}$}
					\\
\hline
{$\Delta M_P$} \\
\cline{0-0}
%\# Resonances \\
${P_1}$&$S_{1\eta}\equiv P_1\oplus\eta_1$ 
 &${\cal R}^c_{20}$&3, 6&28&2.19, 2.27&1&33&15&5&54&79&27&0\\
&&${\cal R}^c_{42}$&3, 8&43&1.90, 1.95&1&99&3&4&6&32&8&15\\
$P'_{1a}$&$P'_{1a}~\oplus~S_{1\eta}$&${\cal R}^c_{20}$&3&29&0.48&4&17&6&3&2&0&25&62&172&126& \\
$P'_{1b}$&$P'_{1b}~\oplus~S_{1\eta}$&${\cal R}^c_{42}$&3, 6&29&2.22&3&128&4&6&6&32&6&9&4&29& \\
%%%%%%%%%%%%%%%%%%%%%%%%%%%%%%%%%%%%%%%%%%%%%%%%
%\\
$P_2$&$S_{2\eta}\equiv P_2\oplus\eta_1$&${\cal R}^c_{10}$&7, 14&3&0.62, 0.80&16&83&44& 19&11&21&173&174       \\
&&${\cal R}^c_{21}$&8, 14&50&0.78, 0.88&3&44&17&4&14&47&37&59\\
&&${\cal R}^c_{31}$&6, 14&93&0.84, 1.04&1&20&18&5&16&43&59&34\\
&&${\cal R}^c_{32}$&13, 19&25&0.8&4&26&4&0&0&3&5&13  \\
&&${\cal R}^c_{43}$&8, 12&72&1.16, 1.26&4&88&13&2&3&12&4&109  \\
%%%%%%%%%%%%%%%%%%%%%%%%%%%%%%%%%%%%%%%%%%%%%%%%
%\\
$P_2$&$S_{2}\equiv P_2\oplus S_{1\eta}$,&${\cal R}^c_{10}$&12&0&0.20&12&144&23& 6&2&3&45&56&90&250&5&216     \\
&$S'_{2}\equiv P'_1\oplus S_{2}$&${\cal R}^c_{21}$&10, 16&97&0.34, 0.50&5&80&23&6&2&3&101&27&83&216&15&185\\
&&${\cal R}^c_{31}$&10, 18&103&0.42, 0.60&6&74&25&3&0&3&14&28&86&224&15&173\\
&&${\cal R}^c_{32}$&15, 19&5&0.82&3&114&31&2&2&11&21&42&146&255&19&213  \\
&&${\cal R}^c_{43}$&10, 14&28&0.84, 0.90&4&46&17&0&0&2&1&6&113&161&15&96 \\
%%%%%%%%%%%%%%%%%%%%%%%%%%%%%%%%%%%%%%%%%%%%%%%%
\cline{0-0}
{\bf $\Delta f_P$}\\
\cline{0-0}
$\eta_1$&--&${\cal L}^c_{-1}$&10, 14&21&0.66, 0.72&2&41&51&3&2&0&\\
%$P_1$&$S_{1\eta}$&${\cal L}^c_{-1}$&12, 14&12&0.30, 032&0&71&62&2&2&0&86&10  \\
$P_1$&&${\cal L}^c_{2}$&4.5, 9&58&1.18, 1.40&0&114&15&1&3&50&9&21&35  \\
$P'_1$&$P'_1~\oplus~S_{1\eta}$&${\cal L}^c_{2}$&7, 9&39&1.38, 1.40&2&139&45&12&14&50&20&67&94&200&8 \\
%&&${\cal L}^c_{3}$&4, 4.5&90&0.72, 0.56&1&49&29&1&1&1&1&88&75&57 \\
$P_2$&$S_{2},~S'_2$&${\cal L}^c_{3}$&13, 18&16&0.58, 0.74&3&32&8&0&0&0&7&0&8&7&2&16&8\\
\\

\hline
%\hline
\end{tabular}
 \caption{{\scriptsize Different sources of errors for the pseudoscalar di-gluonia masses and couplings given in Table\,\ref{tab:res}. 
 %from  LSR for PT at N2LO and gluon condensates at NLO up to $D=8$ and for different parametrizations $S_i$ of the spectral function. 
 The errors from QCD input parameters come from Table\,\ref{tab:param}. The quotd values of $\tau$ for $P_2$ in the cases $S_2$ and  $S'_2$ correspond to $S'_2.$}}
 \label{tab:error}
}
}
\end{center}
\end{table}
%\vspace*{-0.75cm}

 %%%%%%%%%%%%%%%%%%%%%%%%%%%%%%%%%%%%%%%%%%
\section*{Declaration of competing interest}
 %%%%%%%%%%%%%%%%%%%%%%%%%%%%%%%%%%%%%%%%%%
The authors declares that he hasno known competing financial interests or personal relationships that could have appeared to influence the work reported in this paper.
%%%%%%%%%%%%%%%%%%%%%%%%%%%%%%%%%%%%%%%%%

%%%%%%%%%%%%%%%%%%%%%%%%%%%%%%%%%%%%%%%%%%
\section*{Acknowledgements}
%%%%%%%%%%%%%%%%%%%%%%%%%%%%%%%%%%%%%%%%%
Ii is a pleasure thank Graham Shore and Gabriel Veneziano for several exchanges on $M_{\eta_1}$, $\chi'(0)$ and the proton spin in Sections\,\ref{sec:lm2} and \,\ref{sec:spin}. 

 %%%%%%%%%%%%%%%%%%%%%%%%%%%%%%%
% \vfill\eject
 %%%%%%%%%%%%%%%

 \end{document}